%% file: ms.tex
\documentclass[preprint,11pt]{aastex}

\shorttitle{Envelopes Kinematics}
\shortauthors{Tobin et al.}

\newcommand{\mum}{\mbox{$\mu$m}}
\newcommand{\nthp}{\mbox{N$_2$H$^+$}}

\newcommand{\nht}{\mbox{NH$_3$}}
\newcommand{\hcop}{\mbox{HCO$^+$}}

\newcommand{\kmspc}{\mbox{km s$^{-1}$ pc$^{-1}$ }}
\newcommand{\kms}{\mbox{km s$^{-1}$}}
\newcommand{\msun}{\mbox{$M_{\sun}$}}

\begin{document}

\long\def\symbolfootnote[#1]#2{\begingroup%
\def\thefootnote{\fnsymbol{footnote}}\footnote[#1]{#2}\endgroup} 

\title{Complex Structure in Class 0 Protostellar Envelopes III: Velocity Gradients in Non-Axisymmetric Envelopes, Infall or Rotation?\footnotemark[1]}
\author{John J. Tobin\altaffilmark{2,3}, Lee Hartmann\altaffilmark{2},
Edwin Bergin\altaffilmark{2}, Hsin-Fang Chiang\altaffilmark{4}, Leslie W. Looney\altaffilmark{4}, 
Claire J. Chandler\altaffilmark{5}, S\'ebastien Maret\altaffilmark{6}, Fabian Heitsch\altaffilmark{7}}

\begin{abstract}
We present an interferometric kinematic study of morphologically complex protostellar 
envelopes based on observations of the dense gas tracers \nthp\ and \nht. 
The strong asymmetric nature of most envelopes in our 
sample leads us to question the common interpretation of velocity gradients as rotation, given the 
possibility of projection effects in the observed velocities. Several
``idealized'' sources with well-ordered velocity fields and envelope structures
are now analyzed in more detail. We compare the interferometric data to position-velocity
diagrams of kinematic models for spherical rotating collapse and filamentary rotating collapse.
 For this purpose, we developed a filamentary parametrization
of the rotating collapse model to explore the effects of geometric projection
on the observed velocity structures. We find that most envelopes in our sample have PV
structures that can be reproduced by an infalling filamentary envelope projected at different angles within the plane 
of the sky. The infalling filament produces velocity shifts across the envelope that can mimic
rotation, especially when viewed at single-dish resolutions and the 
axisymmetric rotating collapse model does not uniquely describe any dataset. 
Furthermore, if the velocities are assumed to reflect rotation, then the 
inferred centrifugal radii are quite large in most cases, indicating significant 
fragmentation potential or more likely another component to the line-center velocity.
We conclude that ordered velocity gradients cannot be interpreted as rotation alone
when envelopes are non-axisymmetric and that projected infall velocities likely dominate
the velocity field on scales larger than 1000 AU.
\end{abstract}

\footnotetext[1]{Based on observations carried out with the IRAM Plateau de Bure Interferometer, 
Combined Array for Research in Millimeter-wave Astronomy (CARMA), and the NRAO Very Large Array.}
\altaffiltext{2}{Department of Astronomy, University of Michigan, Ann
Arbor, MI 48109}
\altaffiltext{3}{Current Address: Hubble Fellow, National Radio Astronomy Observatory, Charlottesville, VA 22903; jtobin@nrao.edu}
\altaffiltext{4}{Department of Astronomy, University of Illinois at
Champaign/Urbana, Urbana, IL 61801 }
\altaffiltext{5}{National Radio Astronomy Observatory, P.O. Box O, Socorro, NM 87801}
\altaffiltext{6}{UJF-Grenoble 1 / CNRS-INSU, Institut de Plan\'etologie et d'Astrophysique de Grenoble (IPAG) UMR 5274, Grenoble, F-38041, France}
\altaffiltext{7}{Department of Astronomy, University of North Carolina, Chapel Hill, NC}

\section{Introduction}

The angular momentum of protostellar cores and envelopes plays
an important role in star formation.  To the extent that the rotation
of the protostellar cloud is non-negligible, a significant fraction
of the star's final mass must be accreted through a circumstellar disk
\citep{zhu2010,kratter2010}.  In addition, collapse with large amounts of angular
momenta can result in gravitational fragmentation into companion stars 
\citep[][and references therein]{burkert1993,boss1995}. Finally, 
differing initial angular momenta of clouds are likely to produce a distribution of
binary separations and differing initial disk configurations, with implications for planet
formation. Thus, the importance of angular momentum in star and planet formation
process illustrates the need to characterize its magnitude in collapsing protostellar cores.

Observational attempts have been made to characterize the rotation of dense cores and
protostellar envelopes on scales of $0.05 - 0.5$~pc, principally using the dense gas tracers
\nht\ and \nthp\ \citep{goodman1993, caselli2002}. 
Velocity gradients are clearly detected in most objects and observations
with increasing spatial resolution tend to find larger velocity
gradients on the same spatial scales due to less smoothing of the
velocity field. On smaller scales, \citet{volgenau2006} and \citet{chen2007} 
examined the kinematic structures of protostellar envelopes
down to $\sim$1000 AU scales, only finding clear evidence of rotation in a few cases. 
The detected velocity gradients are often interpreted under the assumption of
solid-body rotation, even in protostellar systems; this enables 
simple calculations of angular momentum \citep{goodman1993, 
chen2007, chiang2010, tanner2011}. Several studies have also found the possible signatures of rotation on 
sub-1000 AU scales from high-velocity molecular line wings 
in interferometric data \citep{brinch2007,lee2009}.  

In addition to rotation, the velocity fields of protostellar envelopes will
necessarily include infall motions. Initial models of collapsing protostellar envelopes assumed spherical
geometry \citep{larson1969,shu1977}; later models included angular momentum, which leads
to flattening of the envelope on small scales near the forming disk
\citep{cassen1981,tsc1984}. Infall is typically probed via single-dish spectra, and the blue-shifted asymmetry
of an optically thick line profile (e.g. \hcop\, HCN, and CS) 
is the classic signature of collapse \citep{walker1987,zhou1992,zhou1993}. The inside-out collapse
model \citep{shu1977} is typically adopted to describe the velocity field and coupled with
 radiative transfer modeling to derive infall rates and infall radii 
\citep{zhou1993,wtbuckley2001,narayanan2002}. The derived infall radii are often on the order
of $\sim$5000 AU, indicating that material is likely falling-in from relatively large radii.
Inside the infall radius, the envelope is expected to exhibit a differential rotation
curve, owing to the conservation of angular momentum, the outer core is often assumed
to have solid-body rotation \citep{tsc1984}. However, \citet{wtbuckley2001} noted that one
of their models required inward motions of the envelope out to its boundary, indicating
that inside-out collapse may not be realistic and an ``outside-in'' collapse may be more appropriate
in some cases\citep[e.g.][]{foster1993}.

Contrary to the assumptions of early models, simulations of star formation
in turbulent clouds indicate that infall to protostars is generally
complex, non-axisymmetric, and often filamentary \citep[e.g.][]{bate2003,smith2011}.
Indeed, observations at higher spatial resolution show that most
pre and protostellar cores are elongated \citep{bm1989,myers1991,bacmann2000,stutz2009} and more likely to be
prolate \citep{ryden1996,myers2005}. In the first paper of this
series \citep{tobin2010a}, hereafter Paper I, we used
\textit{Spitzer} IRAC images of a set of Class 0 protostars 
to characterize dense envelope structures in extinction at 8\micron;
the results showed that many systems have very flattened,
filamentary, and often strongly non-axisymmetric density structures.

The recognition of the complexity and non-axisymmetry of many
protostellar envelopes poses substantial problems for any attempt
to infer angular momenta and the interpretation
of line profiles under the assumption of spherical collapse. 
Kinematic observations necessarily refer to integrations along the line of sight of
structures with uncertain geometry, and because only one of
the three velocity components is detected, there is substantial
ambiguity in interpreting the observations as rotation or infall 
\citep[e.g.][Smith et al. 2012 submitted]{dib2010}.  More specifically,
infall (or outflow) could masquerade as rotation, depending upon
the geometry and viewing aspect of the envelopes.

In the second paper of this series \citet{tobin2011}, hereafter Paper II, 
we presented observations of the Class 0
envelopes of Paper I in the dense molecular tracers \nthp\ and \nht.
Using moment maps, we showed that the kinematics of these envelopes,
like their density structures, were often complex and difficult to interpret.
The interferometric data presented in \citet{tobin2011} 
had a higher frequency of envelopes with velocity gradients that \textit{could} 
be interpreted as rotation, as compared to \citet{volgenau2006} and \cite{chen2007};
however, we were not convinced that our data were clearly tracing a rotation signature, even in ideal
cases.

In this paper, we proceed to a more detailed analysis of the kinematics
for a subset of the sample from Papers I \& II where the velocity fields appear to
be rather well-ordered. Mindful of the interpretation difficulties in complex geometry, we develop a simple kinematic model for 
the collapse of a filament to compare with the observations. While significant ambiguities
remain, we find evidence for rotation in some systems on small 
($<$ 1000 AU) scales, but infer that infall motions are likely to dominate the radial
velocity structure on scales $>$ 1000 AU. Our findings emphasize the
importance of following the kinematics to the smallest scales to determine the
angular momenta of infalling envelopes, and the need for further,
more realistic simulations of the large-scale velocity fields in
star-forming regions for comparison with observations.

This paper is
organized as follows: Section 2 briefly describes the models of filament and axisymmetric
collapse, Section 3 describes the comparison of the models to the observed channel maps and PV
 diagrams, and Section 4 discusses our results. The observational details were described
in Paper II and are omitted here for brevity. We will present data for five sources, listed in Table 1, four of
which observed the \nthp\ molecule (one with the Plateau de Bure Interferometer and three with CARMA) and
one source was observed in \nht\ with the Very Large Array\footnote[1]{The National Radio
Astronomy Observatory is a facility of the National Science Foundation operated under
cooperative agreement by Associated Universities, Inc.}.

\section{Kinematic Models}
Most previous work on the velocity structure of protostellar envelopes assume spherical or axisymmetry
in the interpretation of kinematic data. However, in the absence of symmetry, the physical interpretation
of velocity gradients is now non-trivial. The difficulties of interpreting velocity fields of protostellar
envelopes seen in projection are schematically illustrated in Figure 1.
For simplicity we plot only the limiting radial velocities expected
from the emitting envelope.  In the case of spherical symmetry (or axisymmetry in the appropriate
plane), pure uniform rotation (solid-body) directly yields a simple linear velocity gradient across the envelope;
the velocity spread is due to the superposition of velocity components at intermediate scales and disappears
toward the envelope edge. Pure radial infall yields no systematic gradient across the envelope, but an increase in the velocity
dispersion near the central gravitating mass. In general we expect both types of motions to be present,
resulting in the skewed velocity caustics shown in the upper right. Note that a Keplerian rotation
curve is distinct from that of solid-body rotation and appears exactly like the radial
infall curve.

Proceeding to a filamentary envelope, the simplest of non-axisymmetric structures, immediately introduces
a fundamental ambiguity, as both pure solid-body rotation {\em and} pure infall can yield similar velocity gradients
from one side of the envelope to the other (bottom two panels). Including both types of motion
makes the situation even more uncertain. As the observations clearly indicate non-axisymmetric
or filamentary envelopes are common (Paper I), this poses a serious problem for interpretation.
 
In principle, the emission in position-velocity (PV) diagrams, such as indicated schematically here, can
provide some indication of true filamentary or non-axisymmetric structure. For instance, axisymmetric
envelopes (especially those infalling) will tend to have larger velocity dispersions at a given position because material along
the line of sight will have differing velocity vectors than in a highly linear structure.
In addition, material falling-in close to the protostar will tend to yield both red- and blue-shifted
emission at each position, whereas a filament will exhibit more of an abrupt change from red to
blue across the position of the protostar (bottom right panels of Figure 1).  

\subsection{Analytic Model of Filamentary and Axisymmetric Collapse}

Given the complications introduced to the kinematics by non-axisymmetric systems,
we make an initial step toward interpreting the observations
by developing a simple kinematic model for comparison with position-velocity
diagrams. We have modified the rotating collapse model \citep[][hereafter CMU model]{ulrich1976, cassen1981} 
which has been extensively used in studies of protostellar envelopes. 
The infalling gas in the CMU model is assumed to fall in from an initial spherical cloud
rotating with constant angular velocity. With this assumption, the streamlines of the infalling
gas do not intersect, and thus the (supersonic) motion can be considered
in the limit of ballistic trajectories around a central gravitating mass.  
(This neglects the self-gravity of the envelope; but \citet{tsc1984} showed that the CMU 
model could be taken as the inner region of a more general model of self-gravitating spherical 
cloud collapse.)

By neglecting the self-gravity of the infalling gas, we can adopt the axisymmetric velocity field of
the CMU solution. We create a filament by modifying the density distribution such that
all the infalling gas is confined within specified streamlines spanning $\Delta\phi$=30\degr\ in azimuth on opposite
sides of the envelope (mirror symmetry). This filamentary geometry enables us to project the density distribution
of the envelope to different orientations in the sky in order to examine the effects of projection on the 
kinematic structure. The height of the envelope in the z-direction is taken to be comparable to the thickness
of the filament as viewed from top-down, but does not play a crucial role in the observed kinematic structure.

The streamlines comprising the filament correspond to limits in velocity space for emission at a
given position, producing curves which can be compared with observed position-velocity diagrams,
a similar concept to that of using ``caustics'' to analyze position-redshift diagrams of
galaxy clusters \citep[e.g.][]{regos1989}. The velocity components of the infalling, 
rotating gas are given by \citep{ulrich1976}
\begin{equation}
v_r = -\left(\frac{GM}{r}\right)^{1/2} \left(1+\frac{\cos \theta}{\cos \theta_0}\right)^{1/2}
\end{equation}

\begin{equation}
v_{\phi} = \left(\frac{GM}{r}\right)^{1/2} \left(1-\frac{\cos \theta}{\cos \theta_0}\right)^{1/2} \left(\frac{\sin \theta_0}{\sin \theta}\right)
\end{equation}

\begin{equation}
v_{\theta} = \left(\frac{GM}{r}\right)^{1/2} (\cos \theta_0 - \cos \theta) \left(\frac{\cos \theta_0 + \cos \theta}{\cos \theta_0 \sin^2 \theta}\right)^{1/2} ,
\symbolfootnote[1]{\citet{ulrich1976} and \citet{chevalier1983} both have $\sin$ $\theta$ in the denominator,
rather than $\sin^2$$\theta$ as written in \citet{tsc1984} and \citet{hartmann2009}. 
The correct term is $\sin^2$$\theta$; this can be verified by ensuring that 
the squares of the three velocity components add to 2GM/R.}
\end{equation}
describing parabolic motion around a central point mass.
The angle $\theta_0$ is the angle between the orbital plane and the rotation axis, 
while $\theta$ is the angle from the rotation axis to the particle. For simplicity, 
the velocities are only considered for motion nearly in the equatorial plane ($\theta_0$ $\ge$ 89\degr),
making $v_{\theta}$ $\sim$ 0. Note that the term $\cos \theta / \cos \theta_0$, appearing in both $v_r$ and 
$v_{\phi}$ will go to zero as $\theta_0$ $\rightarrow$ $\theta$. Furthermore, most envelopes in
this study are rather filamentary and we will be comparing with PV diagrams taken
from equatorial regions, making $v_{\theta}$ contributions negligible.
A fiducial central object mass of 0.5 $M_{\sun}$ is assumed in these models,
intended to be a typical protostellar mass in the absence of any real
constraints. However, the envelope masses within 10,000 AU are $\sim$1 $M_{\sun}$
for the objects that will be considered in this work (Table 1); this is not negligible 
with respect to the assumed central mass and we will discuss its possible effects
on the kinematics in Section 4.

Other than mass, the only free parameter of the CMU velocity field is the initial 
angular velocity of the material falling in at the current time.
Material falling in from an initial direction given by the angle $\theta_0$ from
the rotation axis lands on the (flat) disk at a radius $r$, given by 
\begin{equation}
\frac{r}{R_C} = \frac{\sin^2 \theta_0}{1-\cos \theta/\cos \theta_0}
\end{equation}
where $R_C$ is the centrifugal radius, 
\begin{equation}
R_C=\frac{R_{0}^4 \Omega^2}{GM}.
\end{equation}
$R_0$ is the radius at which the material fell-in from with angular velocity $\Omega$.
The $\cos \theta / \cos \theta_0$ terms in Equations 1-3 can then be rewritten in terms of $r/R_C$, 
relating the CMU rotation velocity ($v_{\phi}$) to the initial angular velocity ($\Omega$) of the infalling shell. 

To calculate the velocities that would be observed in a PV plot 
across the equatorial plane, the velocities in the $r$, $\phi$,
and $\theta$ directions must be projected along a line of sight. For simplicity, the
line of sight is defined to be the y-axis at z = 0, which makes
\begin{equation}
v_{los} =  v_r (\sin i \sin \Phi_0) + v_{\theta} (\cos i \sin \Phi_0) + v_{\phi} (\cos \Phi_0)
\end{equation}
where $\Phi_0$ is projection angle of the envelope within the line of sight, 
measured counterclockwise from the x-axis. $\Phi_0$ is unrelated to $\phi$ which is
the azimuthal angle of infalling material and $\Phi_0$ only affects
the observed velocity distribution in a non-axisymmetric system.
The angle $i$ is the inclination of the rotation axis of the system with respect to the plane of the sky; this
adds an additional projection term to the observed velocity distribution. However, we 
only consider velocities calculated at inclination of 90\degr, meaning that the rotation axis of the model
would be in the plane of the sky. This simplification is made to limit free parameters and because
most of the observed systems in Paper II have inclinations of $i \geq 60^{\circ}$, making this projection effect
minimal. 

We neglect radiative transfer and chemistry for this initial exploration; this is 
reasonable because we will compare the models to optically thin molecular line kinematics
that trace cold (T$\sim$10K) molecular gas at the densities expected for protostellar envelopes 
($n$ $>$ 10$^5$ cm$^{-3}$, i.e. \nht\ and \nthp).
These molecules are present with detectable emission from large-scales ($\sim$ 0.1 pc)
down to radii as small as $\sim$1000 AU \citep[][Paper II]{caselli2002, bm1989, lee2004}; however, the 
abundances of these molecules are not constant with radius. Our models use the simplest possible
non-axisymmetric envelope structure to describe the high density material and optically
thin emission line kinematics: an m = 2 single filament. The use of optically thick lines to probe the envelope 
kinematics would need to consider lower density surrounding material that could contain even higher-order structure resulting 
in more complex line profiles. Comparison to the numerical models of \citet{smith2011} with radiative transfer
will be the subject of a future paper. 

\subsection{Predicted Kinematic Structure}

The left panel of Figure \ref{axi-streampv} shows the infall streamlines on large scales
for the axisymmetric case, and the middle panels show a zoom-in on the inner 1000 AU. 
The infall streamlines are computed every 15 degrees 
in the $\phi$ direction to sample the entire envelope. 
The projected velocities
versus x-axis position are used to construct simple PV plots 
from the individual envelope streamlines, shown in the right
panel of Figure \ref{axi-streampv}. Note that we have assumed that the envelope 
is infalling from 10,000 AU, with no static outer core. However,
we know that there is substantial mass on larger scales for all these objects measured
via 8\micron\ extinction, continuing to increase from 0.05 pc to 0.15 pc in radius (Paper I).
The masses compared with the observed linewidths indicate that the envelopes are gravitationally
bound (Paper II); therefore, we expect that material from larger scales will be falling in,
 but under the influence of greater enclosed mass. Thus, the large-scale velocities 
from the CMU velocity field are lower limits.

To approximate a filamentary envelope in the same manner as the axisymmetric envelope,
we only calculate three streamlines on either side of the protostar,
spanning $\Delta\phi$=30\degr\ in azimuth at scales larger than $R_C$.
The geometry of the infall streamlines is shown in the top left and middle panels 
of Figure \ref{filament-streampv}. The PV plots are then constructed in the same manner as the
axisymmetric case and shown in the right panels of Figure \ref{filament-streampv}.
We have assumed that matter is infalling from 10,000 AU, the same as the 
axisymmetric case.
Notice that the region of velocity space enclosed by the filament model is smaller than the axisymmetric
case; the velocity streamlines plotted in the filament model represent a subset of those plotted for the
axisymmetric model. Then the effects of viewing the filamentary envelope at $\Phi_0$ = $\pm$30$^{\circ}$
projection angles are shown the middle and bottom rows 
of Figure \ref{filament-streampv}. $\Phi_0$ specifically refers to the projection angle of the middle streamline
on either side of the envelope in Figure \ref{filament-streampv}. Variations of $\Phi_0$ in non-axisymmetric systems
can clearly introduce apparent velocity shifts and gradients that are not present in axisymmetric geometries.

Substantially different velocity structures
are evident in the filamentary PV plots, especially when viewed with different projection
angles within the plane of the sky, as compared to axisymmetric collapse in
Figure \ref{axi-streampv}. Notably the velocity shift across the filamentary envelopes on
scales $>$ 1000 AU is \textit{not} due to rotation velocity; rather it is due to the infall
velocity being projected along the line of sight. \textit{The schematic plots of envelope kinematics
in Figure 1 and the PV plots derived from the CMU model for axisymmetric and filamentary envelopes can be
used a benchmarks to compare with our observational data. This will potentially enable us to distinguish
between axisymmetric and non-axisymmetric infalling envelopes, as well as velocity gradients 
that are due to rotation or projected infall.}

\section{Observational and Model Results}

We have selected five protostellar systems from Paper II to compare with the kinematic models, all having
relatively uncomplicated morphological structure and/or
ordered kinematic structure. These features are important 
because the interpretation of the velocity structures
is complicated by the often complex envelope 
structures (Papers I \& II). We examine channel maps and 
position-velocity (PV) diagrams for each 
protostellar system to better understand the dynamical processes behind the observed kinematic 
structure. We compare the models and data using PV plots of the simplistic 
axisymmetric or filamentary collapse models described in the previous section. The
final model parameters are listed in Table 2.
These models do not provide unique fits, 
given the geometric uncertainties, but they illustrate the basic features
and issues involved in determining the processes giving rise to envelope kinematics.
Also note that we have not combined zero-spacing
data with our interferometric data resulting in the large-scale emission
being filtered out and possibly influencing our kinematic data. However, we show in Paper II that
the velocity fields observed in the single-dish and interferometric data are consistent at large-scales,
indicating that the spatial filtering by the interferometer is not biasing our kinematic results.

\subsection{L1157}
L1157 is located in Cepheus at a distance of $\sim$300 pc and is an example of a 
rather well-ordered, highly flattened envelope.
The flattened envelope is observed in 8\mum\ extinction, as well as \nthp\ and \nht\ 
emission. The \nthp\ integrated intensity map is shown in the upper left panel
of Figure \ref{L1157-chanmaps}. The envelope mass was found to be at least $\sim$0.86 $M_{\sun}$ from 8 \micron\ extinction (Paper I)
and possibly an additional $\sim$0.7 $M_{\sun}$ inferred from the \nthp\ emission (Paper II).
 We have further shown in Paper I that the flattening reflects a filamentary structure in three-dimensions
rather than a sheet extended within the plane of the sky. The scale heights of the short envelope axis
were $\sim$1600 AU across all radii (except the inner 10\arcsec), consistent with
a hydrostatic filament; a sheet would be a factor of three more extended in the vertical direction.
Thus, L1157 may be well-described by an axisymmetric filament,
one of the ideal cases shown in Figure \ref{filscheme}.

The channel maps in Figure \ref{L1157-chanmaps} show several kinematic features in this 
system. In the top and bottom rows, there is higher-velocity \nthp\ emission north and
south of the protostar extended along the outflow. This emission was attributed to
outflow interactions in Paper II. In the middle panels, there are emission peaks
from the inner envelope on $\pm$5\arcsec\ scales present in all the channels. However,
on scales $>$ 7\arcsec\ we see that the east side of the envelope comes into view
at blue-shifted velocities and the west side at red-shifted velocities. This traces a clear
large-scale velocity gradient that may be attributed to rotation or projected
infall. Note that the extension of \nthp\ emission southeast of the protostar traces emission
from the outflow cavity wall and also appears to reflect outflow entrainment.

The PV plot in Figure \ref{L1157-pv-n2hp} also shows high-velocity emission due to the
outflow-envelope interaction extending toward red-shifted velocities on both sides of the envelope,
coincident with the \nthp\ emission peaks. Thus, any indication of 
high-velocity emission due to rotation or infall at small-scales is masked
by the outflow effects. However; \citet{gueth1997} appears to have detected a smaller-scale velocity gradient
in C$^{18}$O emission On scales $>$7\arcsec\ in Figure \ref{L1157-pv-n2hp}, the envelope velocities are
 nearly constant with a systematic $\sim$0.1 \kms\ velocity shift between 
the east and west sides of the envelope, as seen in the channel maps. The larger-scale features, with their nearly
constant velocity, are quite similar to the filament model shown in the bottom right panel of Figure
\ref{filament-streampv}. At the same time, the velocity structure is dissimilar to uniform rotation 
and the axisymmetric infall model \citep[c.f.][]{chiang2010}. A filament model has been overlaid on the data in the PV plot in Figure \ref{L1157-pv-n2hp},
showing that the fit is reasonable, except in the inner envelope where the kinematic structure is not accurately probed. 
The model is projected within the plane of the sky by 
$\Phi_0$ = 15\degr\ and has a centrifugal radius of 100 AU (Table 2).


\subsection{L1165}

L1165 is also located in the Cepheus region and the protostar is forming within a larger-scale
filamentary structure (Paper I). The \nthp\ emission on small scales, shown in 
the upper left panel of Figure \ref{L1165-chanmaps}, traces a structure extended normal to the outflow, peaking just 
southeast of the protostar. A lower limit on the envelope mass is measured to be $\sim$1.1 $M_{\sun}$
from 8\micron\ extinction in Paper I, with perhaps at least an additional $\sim$0.4 $M_{\sun}$ inferred
from \nthp\ emission in Paper II. The \nthp\ channel maps for L1165 are shown
in Figure \ref{L1165-chanmaps}, indicating a clear
velocity gradient across the protostar (normal to the outflow direction). The velocity on the
northwest side of the protostar is blue-shifted and nearly constant. The emission
then becomes red-shifted on the southeast side of the envelope, with the highest velocity emission being red-shifted,
near the protostar. There is evidence of some higher-velocity blue-shifted emission adjacent to the protostar
on the northwest side in the channel maps, but it is not as definitive as the red-shifted emission.

The PV diagram in Figure \ref{L1165-pv-n2hp} shows that the \nthp\ emission
northwest of the protostar has a very narrow linewidth and
a roughly constant velocity; the broadest linewidth is southeast of the protostar, near the \nthp\ peak. 
Southeast of the peak linewidth, the line rapidly becomes narrow again, while becoming more 
blue-shifted.
The channel maps show that the \nthp\ kinematics are unlikely to be outflow
related, due to their location orthogonal to the outflow.
 The PV structure in Figure \ref{L1165-pv-n2hp} is most similar to the filamentary infall
model; its velocities are overlaid in Figure 
\ref{L1165-pv-n2hp}. The model is able to approximate the observed velocity structure,
though there are discrepancies; for L1165 the model is projected by $\Phi_0$=15\degr\ within 
the plane of the sky and has a centrifugal radius of 10 AU (Table 2). The small centrifugal radius
of the model implies that the data are consistent with little rotation on scales 
probed by \nthp.

As Figures \ref{L1165-chanmaps} and \ref{L1165-pv-n2hp} show, \nthp\ is only able to trace emission
to within R$\sim$1200 AU of the protostar. A special feature found in L1165 is that
\hcop\ ($J=1\rightarrow0$) emission is able to trace small-scale kinematics with
higher velocity in the inner envelope, shown in Figure \ref{L1165-pv-hco}. 
The red and blue-shifted components are extended normal to the outflow at radii of 3\arcsec\ (600 AU)
from the protostar. This emission lies inside the broadest \nthp\ emission, suggesting that
it traces smaller-scale kinematic structure, in agreement with the small-scale model velocities
which are overlaid in Figure \ref{L1165-pv-hco}. 
Furthermore, location of the blue and red-shifted high-velocity emission 
on opposing sides of the envelope with opposite velocities, along with
the small velocity width a large-scale in the \nthp\ emission, is consistent
with what is expected for a filamentary envelope (Figure \ref{filament-streampv}).
A similar high-velocity \hcop\ feature was found in RNO43 (Paper II).

\subsection{CB230}

CB230 was classified as a ``one-sided'' envelope 
in Paper I, given its strongly asymmetric distribution of 8\mum\
extinction; this was reflected on large-scales in the single-dish \nthp\ map shown in Paper II.
We measure a lower limit on the envelope mass within 0.05 pc of $\sim$1.1 $M_{\sun}$ in Paper I from 8\micron\
extinction, but there is clearly mass on small-scales that we do not probe. 
The smaller-scale emission probed by \nht, shown in the upper left panel of Figure \ref{CB230-chanmaps}, 
is slightly asymmetric, but well-ordered as a whole. The mass inferred from
the small-scale \nht\ emission in Paper II was $\sim$4.8 $M_{\sun}$, but highly uncertain.
 There is also a depression in the \nht\ emission in the
inner envelope, coincident with the protostar. This effect is attributed to destruction of \nht\ in the inner envelope (Paper II).
The velocity structure of the \nht\ emission is shown by channel
maps in Figure \ref{CB230-chanmaps}. Because the main \nht\ (1,1) lines are a pair separated by 0.2 \kms,
the last three panels in the middle row are blends of blue and red-shifted emission, while the rest of the plots
reflect mostly unblended emission. The channel maps show that there is a clear, well-ordered
velocity gradient across the envelope from east to west, normal to the outflow.

The PV diagram in Figure \ref{CB230-pv-nh3}, in contrast to the channel maps,
is derived from \nht\ satellite emission lines that are separated by $\sim$0.4 \kms.
This makes the emission appear more distinct than if the PV diagram were 
plotted using the main \nht\ lines. The PV plot and channel maps show that 
the emission on the east and west sides of the envelope has a rather constant velocity and
there is an abrupt shift from blue to red-shifted emission starting at the location of the protostar.

While a measure of the small-scale velocities is missing, the emission on
scales $>$1000 AU is most consistent with the infalling
filament, due to the constant velocity emission out to large-scales and the 
relatively narrow linewidth. 
The constant velocity emission is dissimilar from
what would be expected for uniform rotation, which would have linearly increasing
velocity out to large-scales. 
The filament model is overlaid on the data in Figure 
\ref{CB230-pv-nh3}, having projected by $\Phi_0$ = 20\degr\ and has $R_C$ = 10 AU. The model
does not exactly describe the data, but substantial mass in the inner envelope implied by
the \nht\ data could make the velocity shift slightly larger on the scales probed.

\subsection{IRAS 16253-2429}
IRAS 16253-2429 is the most nearby object in our sample, located in Ophiuchus (d$\sim$125 pc) and
one of the most symmetric, spherical envelopes in 8\mum\ extinction
and \nthp\ emission, plotted in Figure \ref{IRAS16253-chanmaps}. The velocity gradient in this 
envelope is quite small, illustrated by the channel maps with the eastern part of the envelope
coming into view just one channel before the rest. Figure 
\ref{IRAS16253-pv-n2hp} shows the PV diagram, the emission
has a roughly constant linewidth and velocity across the envelope and only a slight
velocity gradient is evident. Furthermore, the \nthp\ emission is depressed
toward the protostar, as compared to the surrounding emission. This effect appears to be a depletion/destruction
process since the lines are not optically thick (Paper II).

 This is the only envelope in our
sample that can be reasonably described by an axisymmetric envelope model; however, it was necessary to lower the
mass to 0.1 $M_{\sun}$, such that the linewidth from infall was not too large.
The lower mass is appropriate for this source given its low luminosity 
($L_{bol}$=0.25 $L_{\sun}$; the envelope does have at least 0.8 $M_{\sun}$ of material
within 10,000 AU, with increasing mass on yet larger scales, suggesting that the central object could be more massive.
On the scales probed by the CARMA data (R$\sim$3125 AU; 25\arcsec) the lower-limit on the
mass measured from the 8\micron\ extinction data is just 0.2 $M_{\sun}$. Between 3000 AU and 6000 AU
scales, probed mainly by the single-dish data from Paper II, the linewidth from spherical infall 
with dominant envelope mass would be $\sim$0.2 \kms. Thus, the data for this source appear to be consistent
with spherical infall given the low-mass of the envelope; however, solid-body rotation and filamentary collapse 
cannot be ruled-out by these data. The lack of a significant velocity structure makes 
the kinematic models degenerate for this object.

\subsection{HH211}

HH211 was classified as an irregular envelope in Paper I; 
however,  its structure in \nthp\ appears filamentary, but not as regular as L1157 or
CB230 (Figure \ref{HH211-chanmaps}). The peak \nthp\ emission is just 
southwest of the protostar, coincident with the \nht\ peak \citep{tanner2011}. The 
mass of this envelope is found to be $\sim$1.1 $M_{\sun}$ from 8\micron\ extinction (Paper I)
and $\sim$0.65 $M_{\sun}$ in the inner envelope from \nthp\ emission (Paper II). 
We also find a velocity gradient that is approximately normal to the outflow
and changing.
The PV plot in Figure \ref{HH211-pv-n2hp} shows that to the northeast, the 
velocity gradient is smaller, while southwest it increases,
where is where it blends with an apparent second velocity component
Paper II and \citet{tanner2011}.

On the whole, the data appear most consistent with solid-body rotation,
but this is uncertain given apparent velocity gradient change and the
irregularity of the envelope. The smaller gradient to the northeast 
is consistent with the infalling filament models, which we overlay in
Figure \ref{HH211-pv-n2hp}. The model is projected by $\Phi_0$ = 15\degr\
 and has $R_C$ = 10 AU.The apparent change of the velocity
gradient could be due to complex projection effects or the surrounding 
environment. Even though there is substantial mass measured in this envelope with an 
asymmetric distribution, it is not plausible that it could generate such a large linear velocity
gradient from infall alone in the context of these simple models.
Furthermore, the linewidth of the envelope northeast 
of the protostar is substantially more narrow than the linewidth southwest
of the protostar. At the \nthp\ peak, the linewidth broadens substantially 
and could be related to the outflow as suggested in \citet{tanner2011},
rather than inner envelope kinematics.

\section{Discussion}

The direction of the jet and outflow is thought to correspond
to the rotation axis of the protostar and disk \citep[e.g.][]{shu1987}. 
Since circumstellar disks form in the inner envelope due to conservation of angular
momentum, it has been natural to interpret large-scale velocity gradients which are normal
to the outflow as rotation \citep[e.g.][and Paper II]{goodman1993, caselli2002,
belloche2002, belloche2004, chen2007}. If velocity gradients 
probe rotation, they give a measure of the angular momentum in a collapsing
envelope, enabling the centrifugal radius to be calculated, the radius where a rotationally supported disk will form.
However, it was shown in Paper I that envelopes are often asymmetric and filamentary and in Paper II
that the large-scale kinematics are often quite complex as compared to a linear velocity gradient.
These factors, along with the examples shown in Figure \ref{filscheme},
make it uncertain whether or not velocity gradients in these envelopes reflect rotation alone.
The qualitative arguments shown in Figure \ref{filscheme} are reinforced by the simple models of axisymmetric
and filamentary collapse shown in Figure \ref{axi-streampv} \& \ref{filament-streampv}.

One of the keys to disentangling the kinematics in protostellar envelopes is higher
resolution data provided by interferometers. This is because rotation velocities are
expected to become most important on small-scales due to conservation of angular momentum. 
Thus, even if the velocity structure on larger-scales probed by single-dish observations
is not wholly due to rotation, we would naively expect rotation velocities to 
become more important on smaller scales.
The interferometric PV diagrams and channel maps of these systems 
reveal small-scale kinematic detail invisible in single-dish data,
but are also sensitive to emission on size scales between 1000 and 10000 AU.
On this scale, differences between the various model PV diagrams 
shown in Figures \ref{axi-streampv} and \ref{filament-streampv} become apparent.
The sources presented in this paper represent some of the most well-ordered systems kinematically and
morphologically from Paper II. However, in spite of this, there are still 
difficulties in disentangling rotation versus infall in the velocity fields and PV diagrams
which must be recognized when using data of this kind to test theories of star formation.


\subsection{Velocity Gradients as Rotation}

It was shown in Paper II that the envelopes often have substantial velocity
gradients ($\sim$2.3 \kmspc) out to 10000 AU and beyond, measured by single-dish mapping.
The majority of the envelopes in the sample also had velocity gradients that
were within 45\degr\ of normal to the outflow direction, reinforcing an interpretation as rotation. 
Previous work has often made the assumption that the envelope velocity
profile reflects solid-body rotation, even at high-resolution \citep[e.g.][]{chen2007}. The protostars
L1165, CB230, and L1157 are inconsistent with solid-body rotation with
their constant velocities out to large-scales. Therefore, if the line-center velocities reflect rotation, then it must
be a differential rotation curve with non-constant angular momentum; constant angular momentum
would yield v $\propto$ R$^{-1}$. HH211 specifically could be
consistent with solid-body rotation and IRAS 16253-2429 could also be represented by solid-body rotation but not
uniquely. If the characteristic large-scale velocity gradient for HH211 from Paper II (6.9 \kmspc) is interpreted as rotation,
the inferred centrifugal radii of this material is $\sim$2800 AU by applying Equation 5.
This assumes that material from 10000 AU fell-in to
to $R_C$, with a 0.5 \msun\ central object. 
However, this calculation assumes a constant angular velocity for the 
entire envelope.

With the interferometer and single-dish data from Paper II, we can measure
the line-center velocity from the inner to outer envelope at many positions and calculate
the centrifugal radius of material throughout the envelope.
We have calculated $R_C$ versus radius from the observed line-center velocities, 
assuming a central object mass of 0.5\msun\ and that the observed velocities reflect rotation. The
results in Figure \ref{RCR} show that inside of $\sim$5000 AU the centrifugal radii are below
1000 AU. However, on scales greater than 5000 AU, the centrifugal radii are in excess of 1000 AU,
except for IRAS 16253-2429. HH211 notably has an unrealistic $R_C$ $>$ 10000 AU at $R$ $\sim$ 6500 AU.
All these centrifugal radii are substantially larger than the characteristic
sizes of circumstellar disks ($\sim$250 AU \citep{andrews2007,andrews2009}). We note
that $R_C$ could be smaller, if the central object masses are larger than assumed. Moreover,
if much of the material observed on 10000 AU scales does become incorporated 
into the protostar, then the implied centrifugal radii will be smaller by about a factor of 3 at $R$ = 10000 AU 
since the masses at $R$ $<$ 10000 AU $\sim$ 1 $M_{\sun}$; however, this does not fully alleviate the problem
since corrections will be less at smaller radii.

Rotation of the measured magnitude is expected to cause fragmentation during collapse 
on large, $\sim$1000 AU, scales \citep[][Zhu et al. 2011]{rafikov2005,rafikov2007,kratter2010}.
Of the sources shown, only CB230 is a wide binary with a separation of
$\sim$3000 AU. However, the gas kinematics around CB230
do not indicate rotationally supported motion on this scale, suggesting that the companion did not form via
rotational fragmentation. The large centrifugal radii inferred for four of five sources, especially HH211,
may indicate that another dynamical process is contributing to the observed line-center velocities.
If the observed velocities are indeed rotation, then to prevent fragmentation, the outer envelopes must be 
removed before they fall-in to the centrifugal radius. Outflows are one way to entrain and evacuate ambient
material \citep{arce2006}; however, it seems unlikely that this mechanism can efficiently work for 
highly filamentary envelopes that are extended in the direction normal to the outflow. 
The physical implications of the observed velocities reflecting only rotation seem to suggest
that there are additional contributions to the velocity field from dynamical processes other than
rotation.

\subsection{Velocity Gradients from Projected Infall}

Since all these systems appear to be gravitationally bound (Paper II), the most likely
process that could contribute to the line-center velocity field
away from the region of outflow influence is infall. Figure 
\ref{filscheme} shows that if an envelope is not axisymmetric, then 
infall could manifest itself in the observed velocity field across the envelope and 
a velocity gradient normal to the outflow does not necessarily imply rotation, also see Figure \ref{filament-streampv}.
The models demonstrate that the radial motion of a globally infalling
filamentary envelope can produce detectable velocity shifts on either
side of the envelope even with only $\Phi_0$ = $\pm$15\degr\ projection
angle within the plane of the sky. The velocity shift remains relatively constant on the same scales
that we see in our observations. Moreover, if the envelope is only marginally resolved (as in the
single-dish observations), such a velocity shift would likely present itself as a linear gradient
in lower-resolution single-dish data.

CB230, L1157, and L1165 all have 
velocities that are approximately constant between R$\sim$1500 and 10000 AU in the
PV plots (Figures \ref{L1157-pv-n2hp}, \ref{L1165-pv-n2hp}, \& \ref{CB230-pv-nh3}; Table 2). 
The PV structures agree with the predicted kinematic structure for an infalling 
filament and are inconsistent with what would be expected from axisymmetric collapse or solid-body rotation.
The resemblance of the PV plots to the predictions of the infalling filament model, as well as the 
filamentary morphology of the envelopes, leads us to suggest that the principal dynamic process producing
the velocity gradients is large-scale infall and not rotation.
However, IRAS 16253-2429 could be described by
any of the kinematic models and HH211 appears most
similar to rapid solid-body rotation but the non-constant gradient and overly large centrifugal
radii may indicate that something more complex is taking place.

The results from the analysis of these data lead us to conclude that many 
of the large-scale velocity gradients observed in Paper II and
other studies most likely arise primarily from infall velocities being projected along our line of sight,
with rotation being a subordinate contribution to the velocity field.
Furthermore, we suggest that even systems which appear to have linear
velocity gradients on large-scales probably do not reflect rotation, given the results for HH211. The linear
velocity gradients probably arise from complex projection effects and the formation of cores in
a turbulent medium.

Evidence for large-scale infall has been previously seen in the star-less core L1544 \citep{tafalla1998} 
and possibly NGC 1333 IRAS 4B \citep{difrancesco2001,jorgensen2007}. Moreover, radiative transfer
modeling of the blue-asymmetric line profiles of optically thick tracers indicates
that infall radii are often of order 5000 AU \citep{walker1987,zhou1993,wtbuckley2001,narayanan2002}. Furthermore, starless cores
forming in the filamentary dark cloud L1517 appear to have material flowing toward the dense
cores \citet{hacar2011} as viewed in the line-center velocities of the
filaments. The large-scale flow of material towards the star-less cores embedded
within filaments is consistent with our picture of the envelopes infalling
from large-scales in these protostellar systems. We may be seeing the later evolution of this
process in our sample, while \citet{hacar2011} is seeing the beginning.

\textit{The possibility of projected infall makes the true rotation velocities difficult to
disentangle in protostellar envelopes with complex morphological structure,
limiting the ability to definitively measure angular momentum at large-scales.} Rotation
must be measured at smaller-scales, where the rotation velocity will be larger due
to conservation of angular momentum during collapse.

\subsection{Effects of Extended Mass}

The CMU infall model assumes a dominant central mass
in the calculation of the velocity field. However, there is substantial
mass on scales larger than $\sim$1000 AU in all sources that could certainly affect the kinematics
observed. We can simplistically characterize the effects of extended mass by employing
Gauss's law for gravity, in which any mass interior to it will appear as a point mass and the
free-fall velocity is
\begin{equation}
v_{ff} = \left(\frac{2GM_{enc}}{R}\right)^{1/2}.
\end{equation}
Since M $\propto$ $\rho(R)$R$^3$, we can rewrite the free-fall velocity with the following proportionality
\begin{equation}
v_{ff} \propto \rho(R)^{1/2}R.
\end{equation}
Thus, an object with a spherically averaged density profile of $\rho(R)$ $\propto$ $R^{-2}$ will have $v_{ff}$ $\sim$ constant
out to large-scales. Most of the objects examined in this paper have approximately constant velocities on large-scales, 
strongly suggesting that they are infalling and extended mass is influencing their kinematics. Even with significant
extended mass, the predictions of the filament model remain valid;
extended mass simply increases the projected infall velocities on large-scales, such that they do not approach
zero as quickly. In the axisymmetric case, the extended mass would make the infall velocities remain quite
broad on large scales, inconsistent with the observations. Lastly, projected infall could mimic
the signature of solid-body rotation if the density were constant with increasing radius; however, on the scales of the 
protostellar envelopes ($\sim$0.1 pc) there does not appear to be a region of constant density in the molecular line emission or
8\micron\ extinction. Nevertheless, this effect could be evident in larger-scale molecular clouds 
such as those observed by \citet{arquilla1986}.

\subsection{Linewidth Constraints on Infall and Morphology}

Evidence for large-scale infall is also found in the linewidths of the envelopes. 
The \nthp\ and \nht\ linewidths in protostellar envelopes are often observed to be about a factor of
two or three larger than expected for purely thermal linewidths 
(0.13 \kms\ for \nthp\ and 0.22 \kms\ for \nht) for T = 10 K gas \citep[e.g.][Paper II]{caselli2002, chen2007}.
Infall motions themselves on large-scales, in an axisymmetric envelope,
would cause substantial line broadening ($\sim$0.5 \kms) due to the superposition
of radially infalling material along the line of sight (Figure \ref{axi-streampv}).
The line broadening generated by the infalling motions of filamentary
envelopes in Figure \ref{filament-streampv} is often just 0.2 \kms\ between radii of 2000 AU and
10000 AU, which is consistent with what is observed in the PV plots and in the linewidth fields
presented in Paper II. Thus, large-scale infall 
of axisymmetric or non-axisymmetric protostellar envelopes
could naturally give rise to the observed line broadening and it would not be necessary to 
invoke a turbulent velocity component to explain the observed linewidths. 

\subsection{Comparison with Simulations}

Simulations of protostellar core/envelope formation within large-scale molecular clouds
also support the idea that rotation is not being accurately probed by line-center
velocity maps. \citep{burkert2000} showed that for individual cores, the 
line-center velocity gradient cannot be assumed to probe the angular momentum for individual
cores, only the distribution of angular momenta can be derived. More recently,
\citet{dib2010} found that angular momenta from cores estimated from the line-center
velocity field versus 3D velocities are overestimated by a factor of $\sim$10. However, they only remark that the difference
arises from the line-center method fitting a global linear gradient and the 3D angular momentum 
including contributions from ``complex dynamical behavior,'' which
could be taken to mean turbulence and/or infall. Finally, \citet{smith2011} investigated the collapse
of filamentary cores in their global simulation, finding that there is infall from larger-scales
toward the cores themselves. Furthermore, recent detailed analysis of the velocity fields show significant
complexity with contributions from infall, turbulence, and some net rotation (Smith et al. 2011 in preparation).
Realistic simulations of collapse have the promise to enable a better understanding of the net angular momentum in cores
as they will enable statistical comparisons to observational data, but with the ability to differentiate rotation and
infall.

\section{Summary and Conclusions}

We have presented an analysis of the dynamical processes giving rise
to the observed kinematic structure in protostellar envelopes. This was done
by comparing interferometric \nthp\ and \nht\ channel maps and position-velocity diagrams
to our qualitative expectations for envelope kinematics and the predicted kinematics of collapse models.
We have constructed an analytic model that approximates filamentary
infall, based on the rotating collapse model \citep{ulrich1976,cassen1981}, enabling
us to explore projection effects in a simple non-axisymmetric system.
We have matched these models to the observations in order to determine
the likelihood of rotation or infall in an axisymmetric or filamentary envelope producing
the observed velocity structure.
The PV diagrams of the infalling axisymmetric and filamentary models
are generally distinct from each other on scales between 1000 and 10000 AU.
The models of filamentary collapse viewed at different projection angles show that
infall motions along the long-axis of a protostellar envelope can produce systematic velocity
shifts across the protostar without the need for significant rotation at large-scales.
These velocity shifts when viewed
at low resolution in single-dish measurements would give the appearance of an approximately linear velocity
gradient.

The velocity structures of L1157, CB230 and L1165 are found to be
inconsistent with uniform rotation and tend to have
constant velocity outside the inner envelope. If the envelope velocities reflected rotation,
then the centrifugal radii calculated on scales larger than 5000 AU would be $>$ 1000 AU, larger
than a typical circumstellar disk. 
The envelope of HH211 has a velocity structure consistent with solid-body rotation; however, the
envelope is a complex filament and the implied centrifugal radii are unrealistically
large. IRAS 16253-2429 has the most symmetric, spherical envelope in the 
sample and also has very little velocity structure, making all the kinematic models degenerate; however,
if rotation is assumed the implied centrifugal radii are reasonable for this object.

Given that the kinematics of L1157, CB230 and L1165 seem to reflect projected infall and HH211
likely has some infall component to its velocities, we therefore suggest the following scenario.
The protostellar envelopes we have observed are in a state of global collapse, and the projection
of the infall velocities along the filamentary envelopes gives rise to the velocity
distributions observed in the PV plots and moment maps in Paper II. This suggests that envelopes with greater
degrees of morphological complexity should have similarly complex velocity fields resulting
from projected infall velocities. 

While there must certainly be net rotation in these envelopes, the degree
to which it is solid-body is uncertain and global simulations
indicate that the turbulent medium from which the cores formed
would impart net angular momentum to the envelope. While we cannot unequivocally prove that
our observations are probing infall rather than rotation, we have shown that it is a viable 
interpretation in asymmetric systems. Moreover, the possibility of infall velocity components
in line-center velocity fields indicates that measurements of angular momentum
on large-scales can be highly inaccurate. To more definitively measure angular
momentum, one must look on small-scales where conservation
of angular momentum may make the rotation velocities larger and more apparent.

We thank the anonymous referee for a thoughtful review, improving the clarity of the final manuscript. The
authors also wish to thank A. Goodman, R. Smith, S. Offner, and M. Borkin for useful discussions. 
We wish to thank the CARMA observers for carrying out the observations. 
We thank V. Pi\'etu for assistance with the PdBI data reduction.
Support for CARMA construction was derived from the states of Illinois, California,
and Maryland, the James S. McDonnell Foundation, the Gordon and Betty Moore Foundation,
the Kenneth T. and Eileen L. Norris Foundation, the University of Chicago, the Associates of the
California Institute of Technology, and the National Science Foundation. Ongoing CARMA development
and operations are supported by the National Science Foundation under a cooperative agreement,
and by the CARMA partner universities. The National Radio Astronomy Observatory is a facility of
the National Science Foundation operated under cooperative agreement by Associated Universities, Inc. 
IRAM is supported by INSU/CNRS (France), MPG (Germany) and IGN (Spain). J. J. Tobin acknowledges support
from HST-GO-11548.04-A, the University of Michigan Rackham Dissertation Fellowship, \textit{Spitzer} archival research program 50668,
and Hubble Fellowship grant
HF-51300.01-A awarded by the Space Telescope Science Institute, which is operated by the Association of Universities for
Research in Astronomy, Inc., for NASA, under contract NAS 5-26555.

{\it Facilities:}  \facility{CARMA}, \facility{VLA}, \facility{IRAM:Interferometer}, \facility{Spitzer (IRAC)}

\begin{small}
\bibliographystyle{apj}
\bibliography{ms}

\clearpage

\begin{figure}
\begin{center}
\includegraphics[scale=0.5]{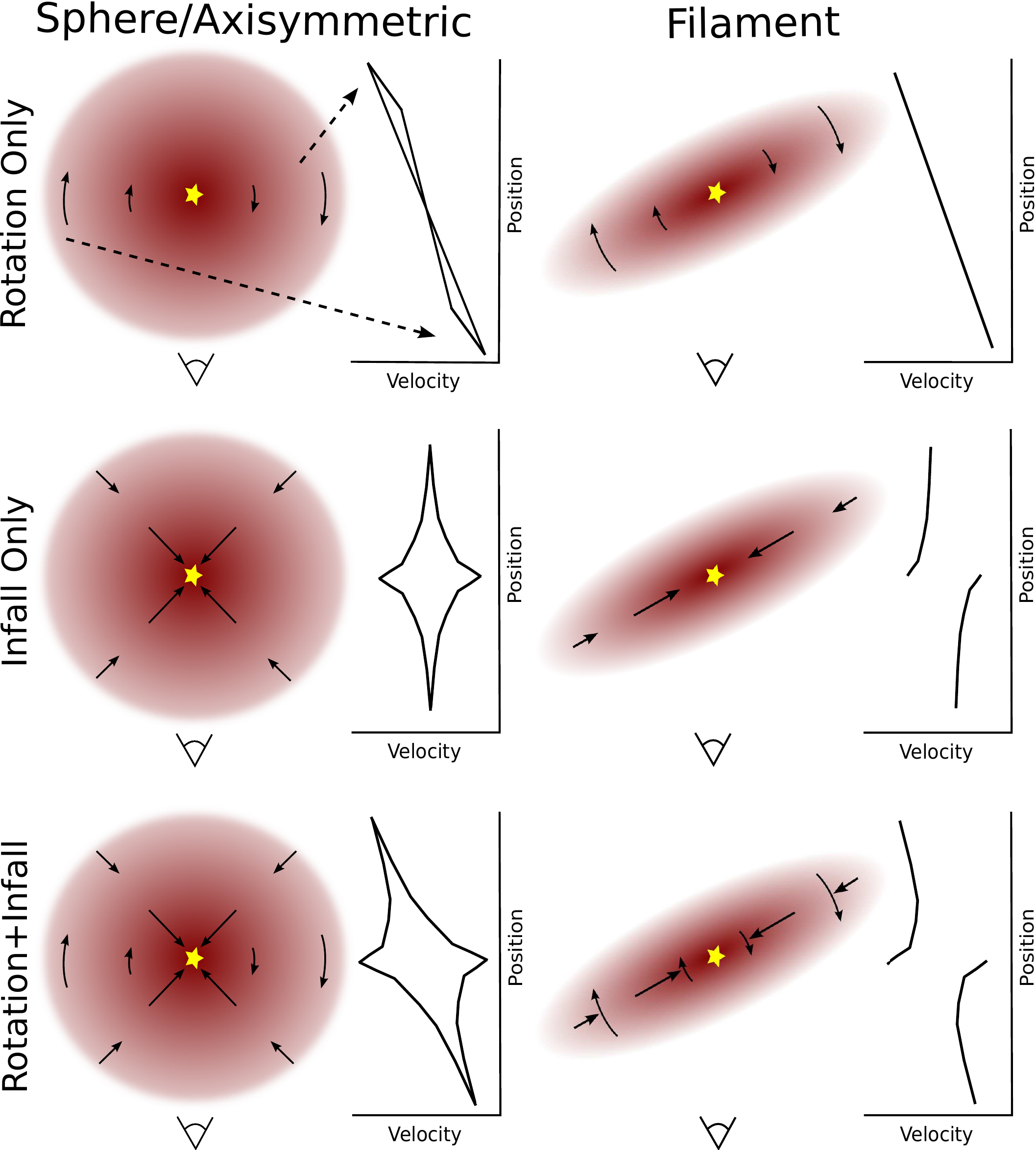}
\end{center}
\caption[Schematic representation of expected velocity structures in envelopes]{These plots schematically show the velocity structure expected to be observed from the dynamical
processes of rotation, infall, and their combination for particular envelope morphologies.
The top row demonstrates spherical/axisymmetric
envelopes and the bottom row demonstrates filamentary envelopes, with schematic position-velocity diagrams below each drawing.
The plots illustrate that solid-body rotation in an axisymmetric envelope (\textit{top left}) 
will show a linear velocity gradient and infall
alone (\textit{top center}) only increases linewidth. The convolution of these processes is shown in
the \textit{top right} panel. A filament
also shows a linear velocity gradient for uniform rotation (\textit{bottom left}); however, infall along a filament
turned toward our line of sight would generate an approximately linear velocity structure in the absence
of rotation (\textit{bottom middle}). When these process are convolved in the \textit{bottom right} panel,
the two processes are difficult to separate.
}
\label{filscheme}
\end{figure}

\clearpage

\begin{figure}
\begin{center}
\includegraphics[scale=0.3]{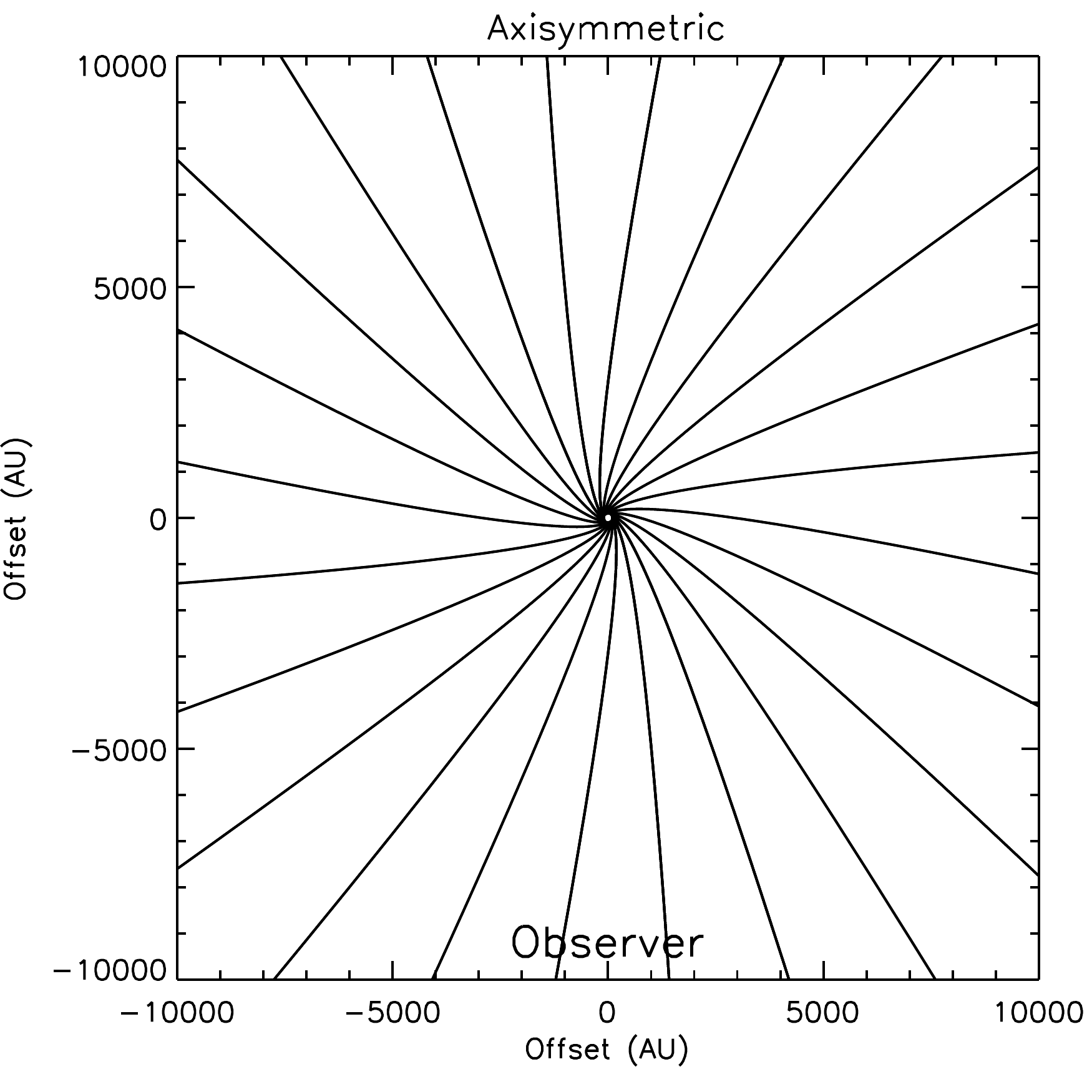}
\includegraphics[scale=0.3]{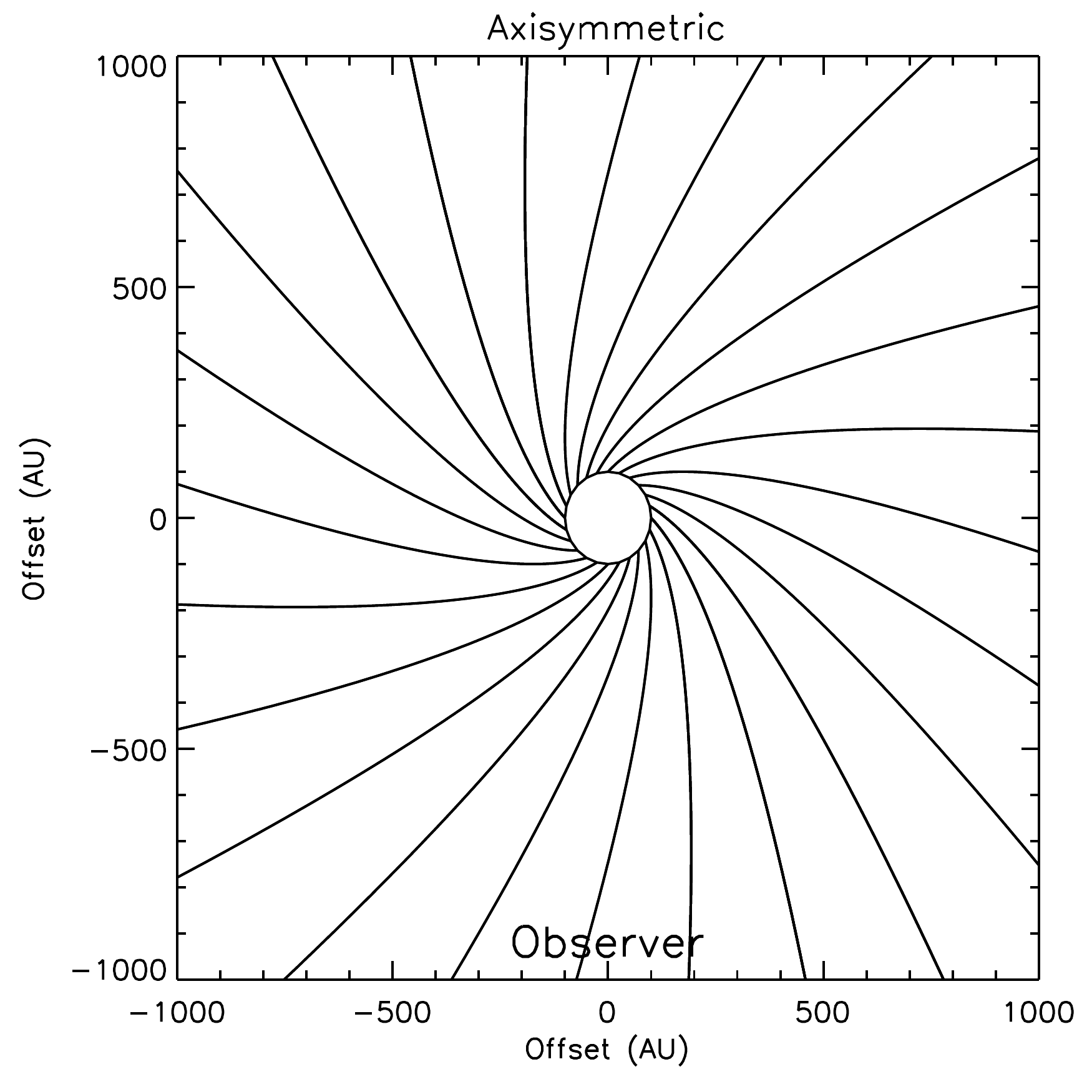}
\includegraphics[scale=0.3]{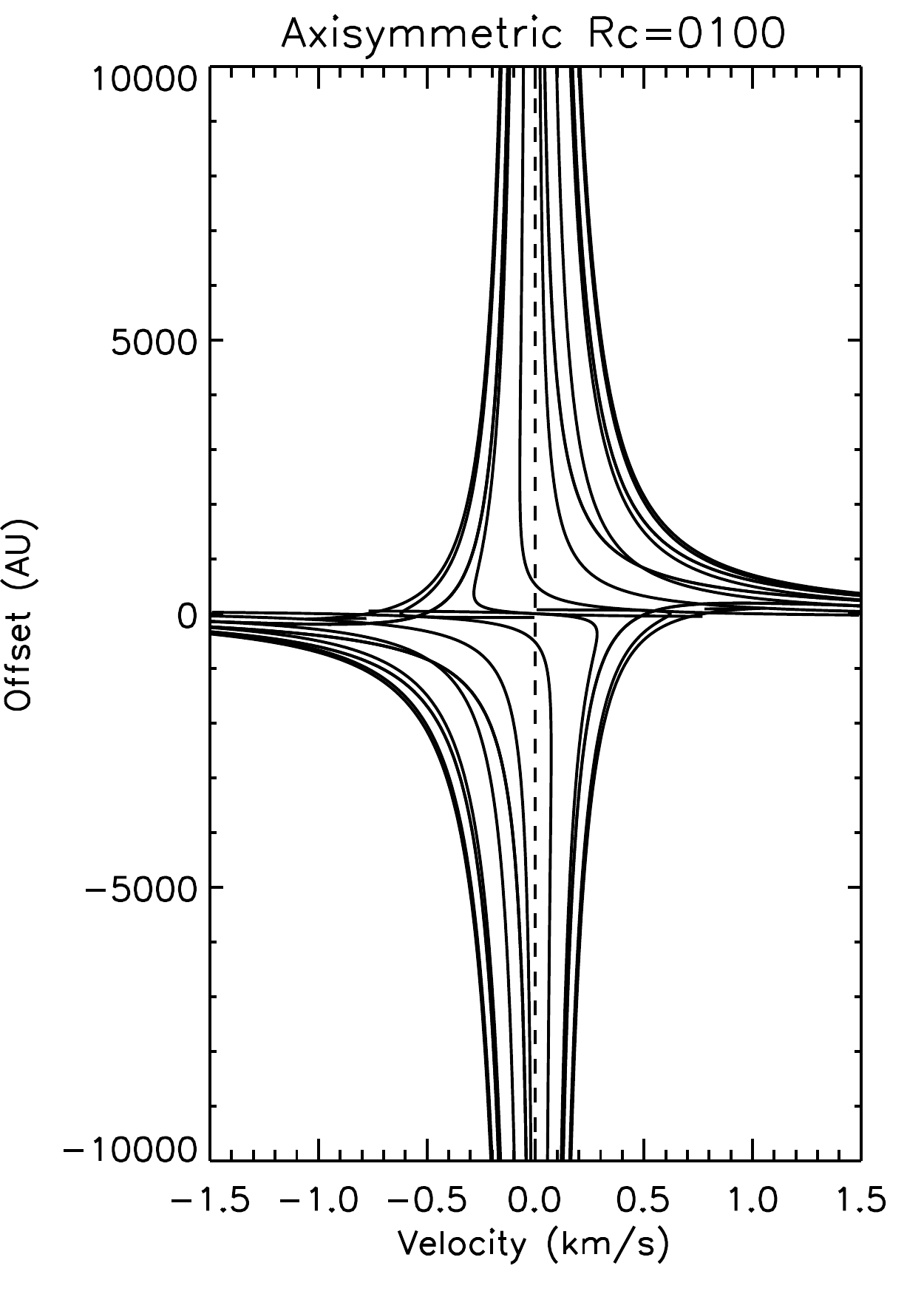}
\end{center}
\caption{Plots of infall streamlines and projected velocities for an axisymmetric envelope 
with $R_C$=100 AU. The \textit{left panel} shows the streamlines
out to 10000 AU while the \textit{middle panel} zooms in on the inner 1000 AU. The circle at the center represents the 
edge of the circumstellar disk forming at $R_C$. The right panel shows the predicted PV structure for the axisymmetric
envelope.
}
\label{axi-streampv}
\end{figure}
\clearpage

\begin{figure}
\begin{center}
\includegraphics[scale=0.3]{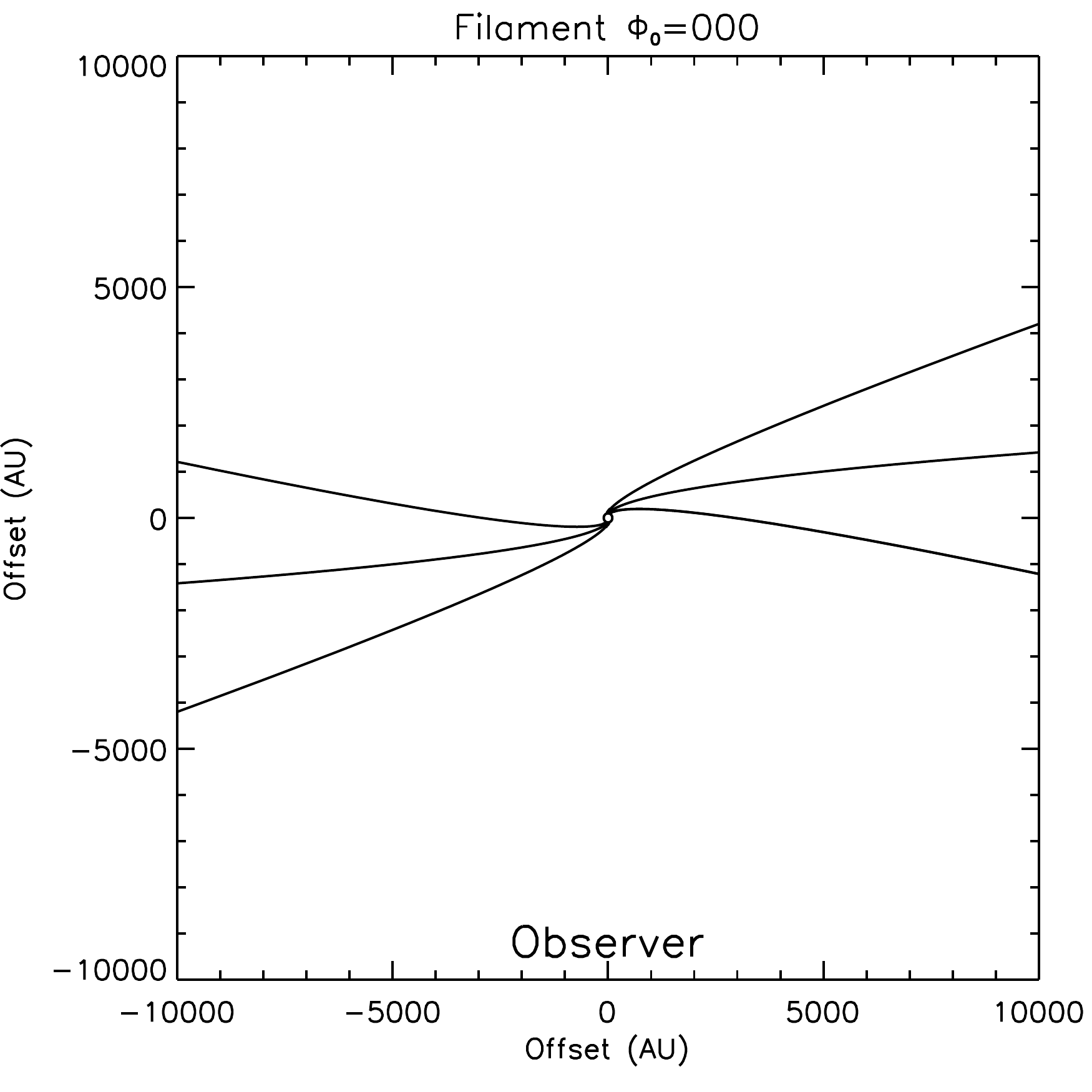}
\includegraphics[scale=0.3]{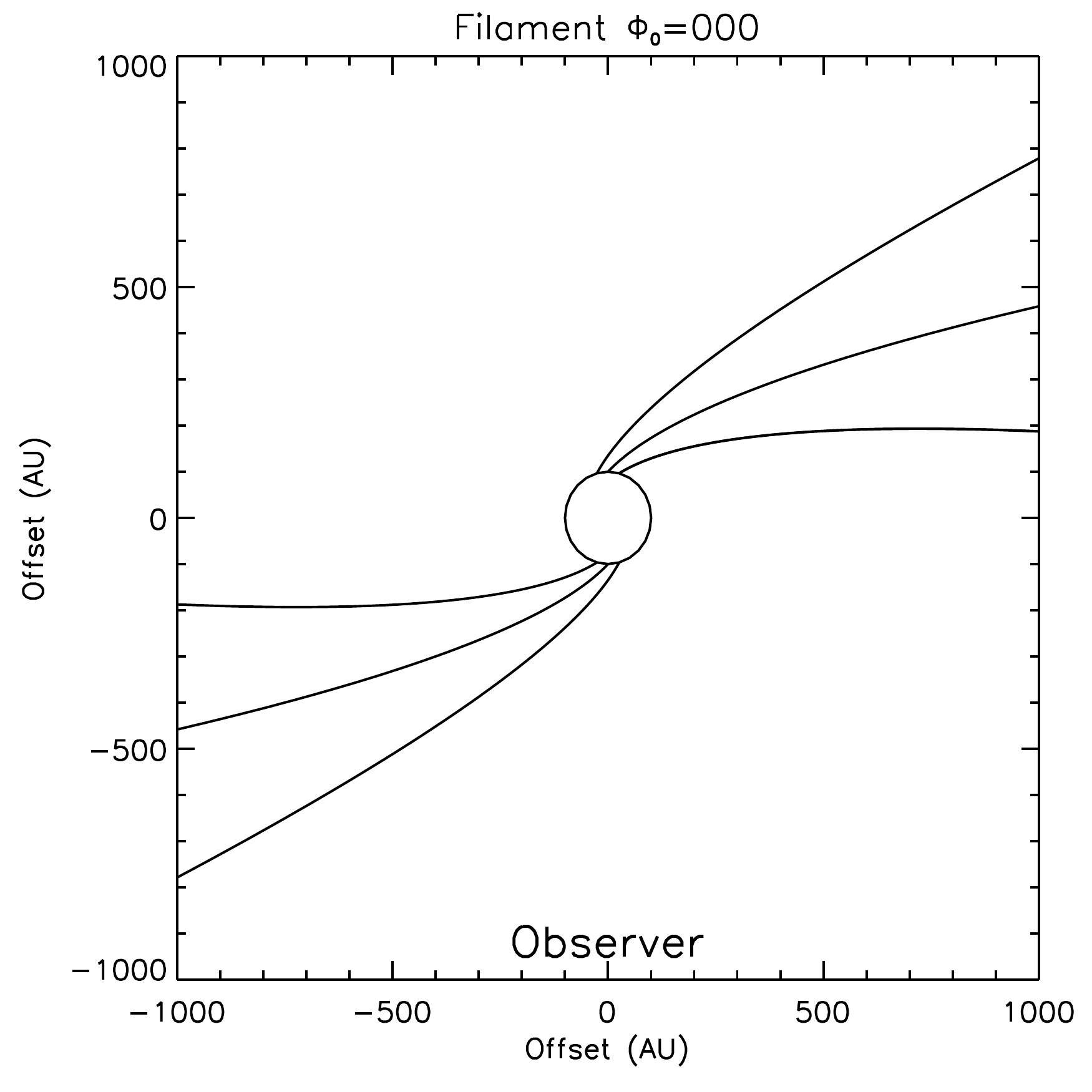}
\includegraphics[scale=0.3]{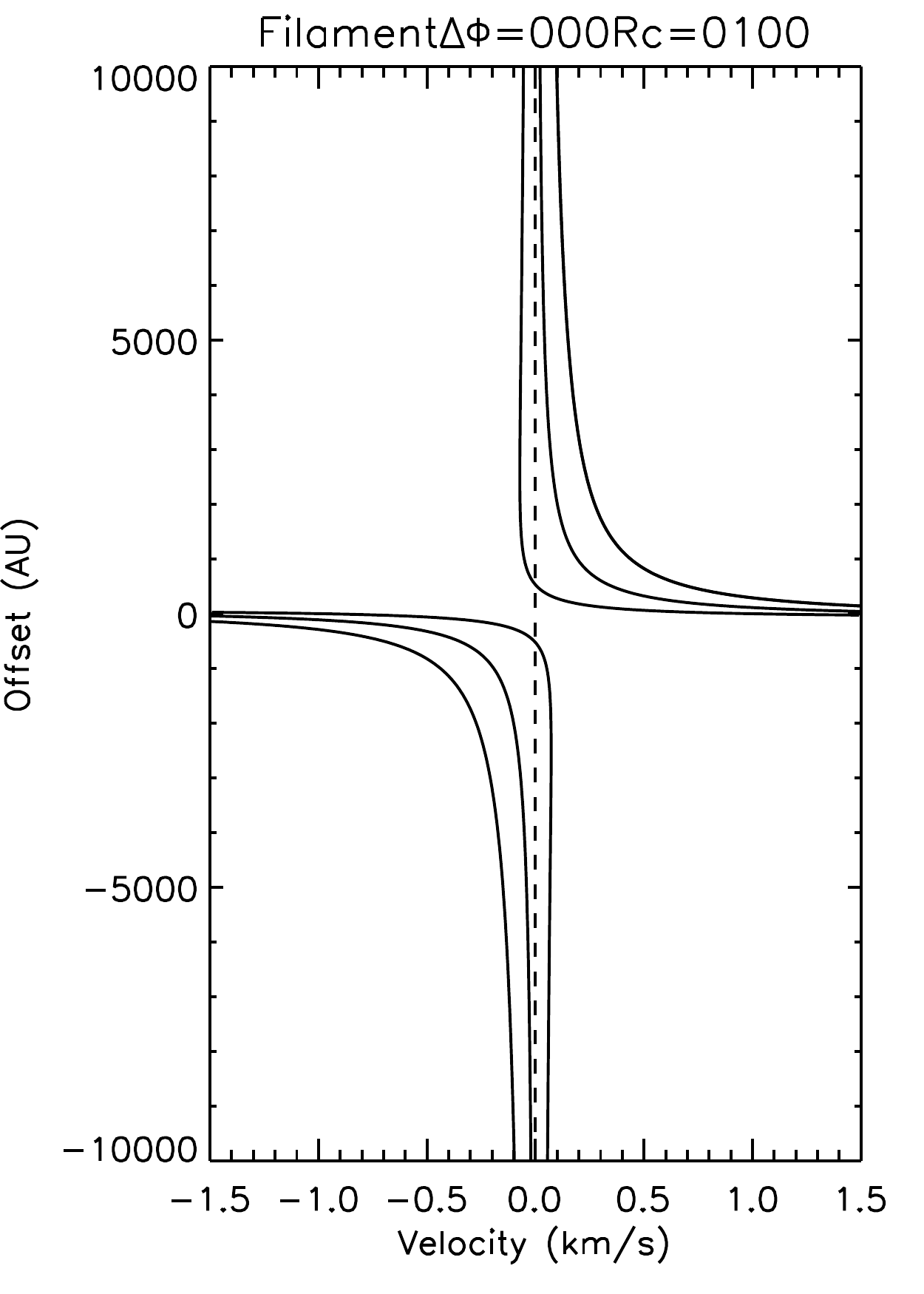}
\includegraphics[scale=0.3]{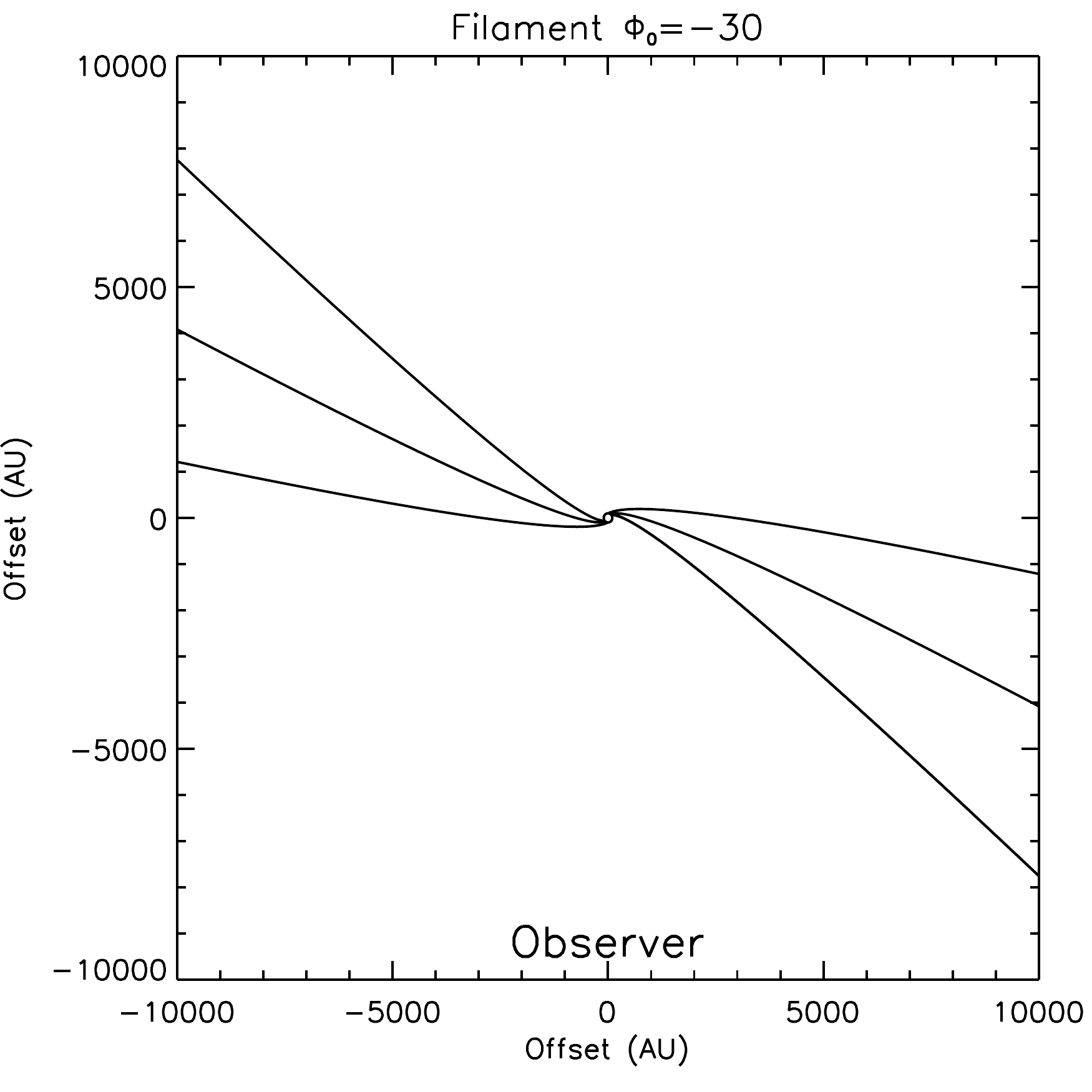}
\includegraphics[scale=0.3]{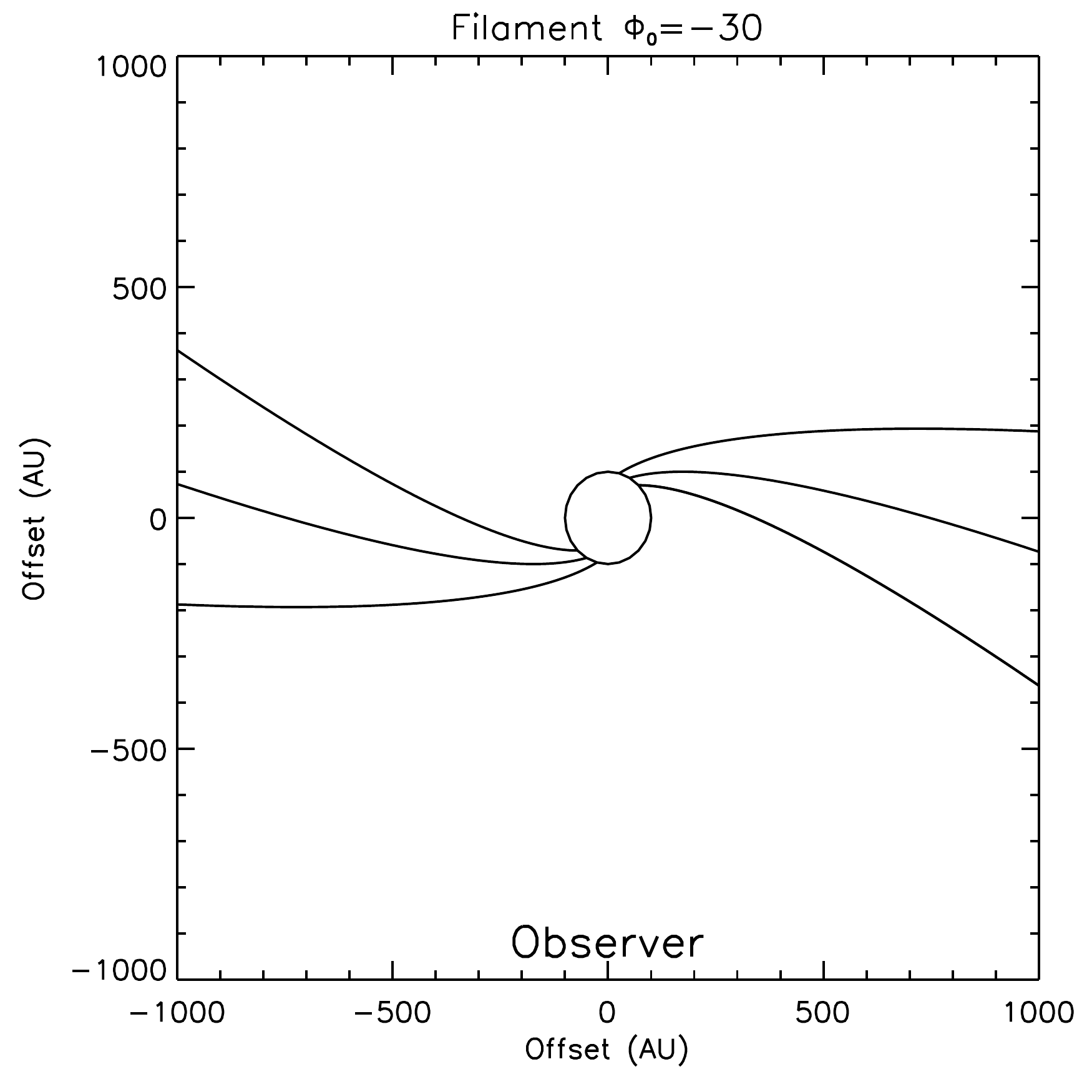}
\includegraphics[scale=0.3]{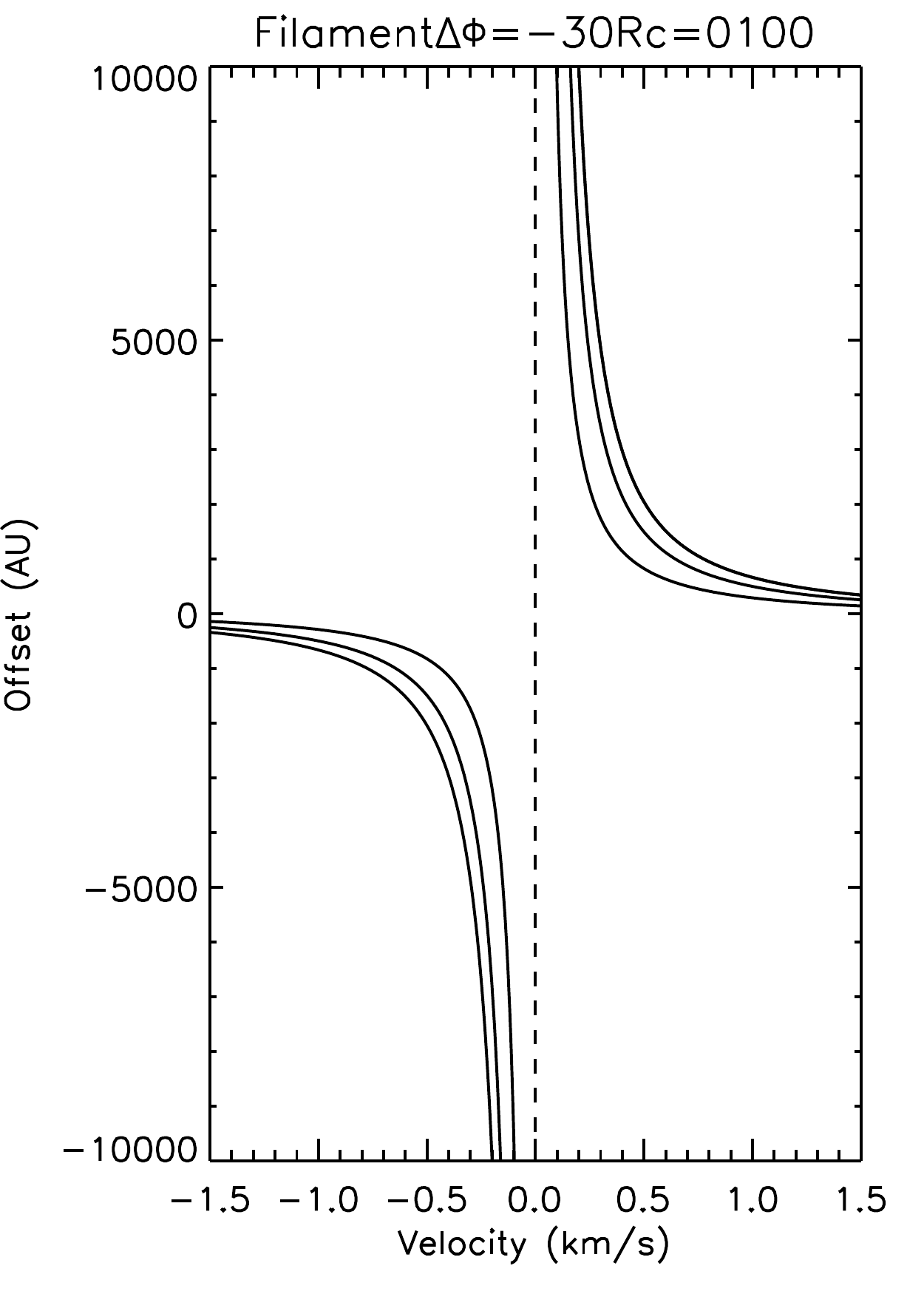}
\includegraphics[scale=0.3]{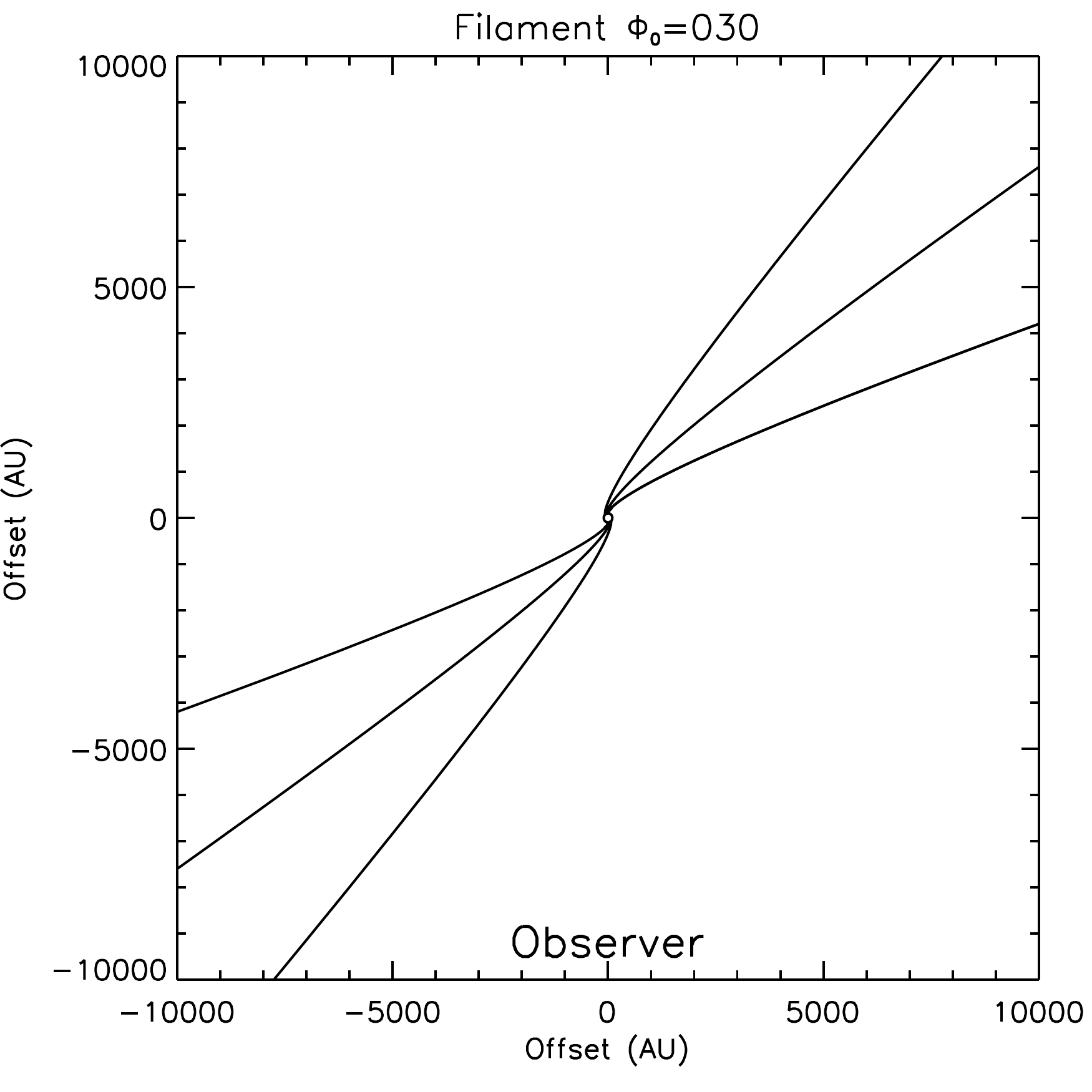}
\includegraphics[scale=0.3]{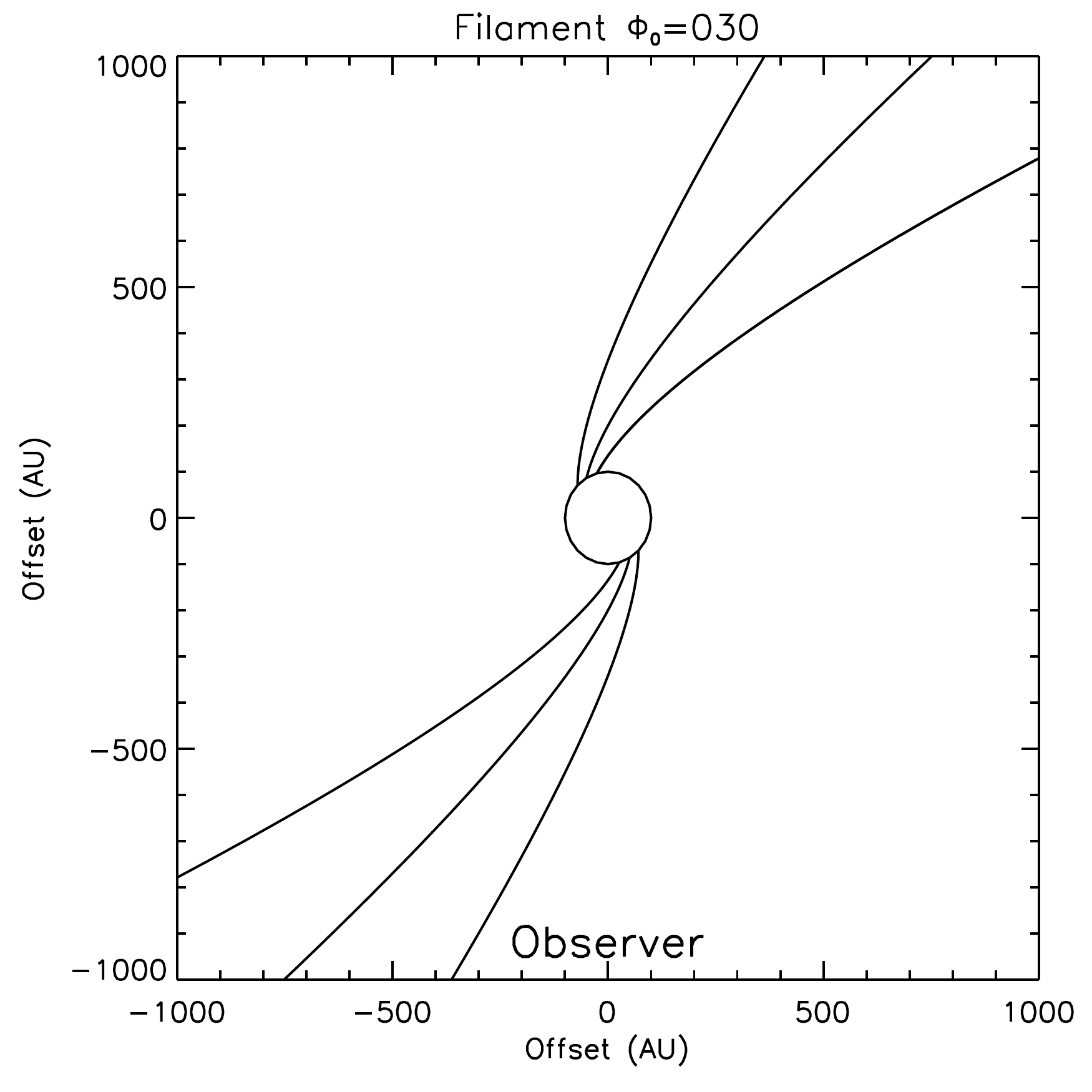}
\includegraphics[scale=0.3]{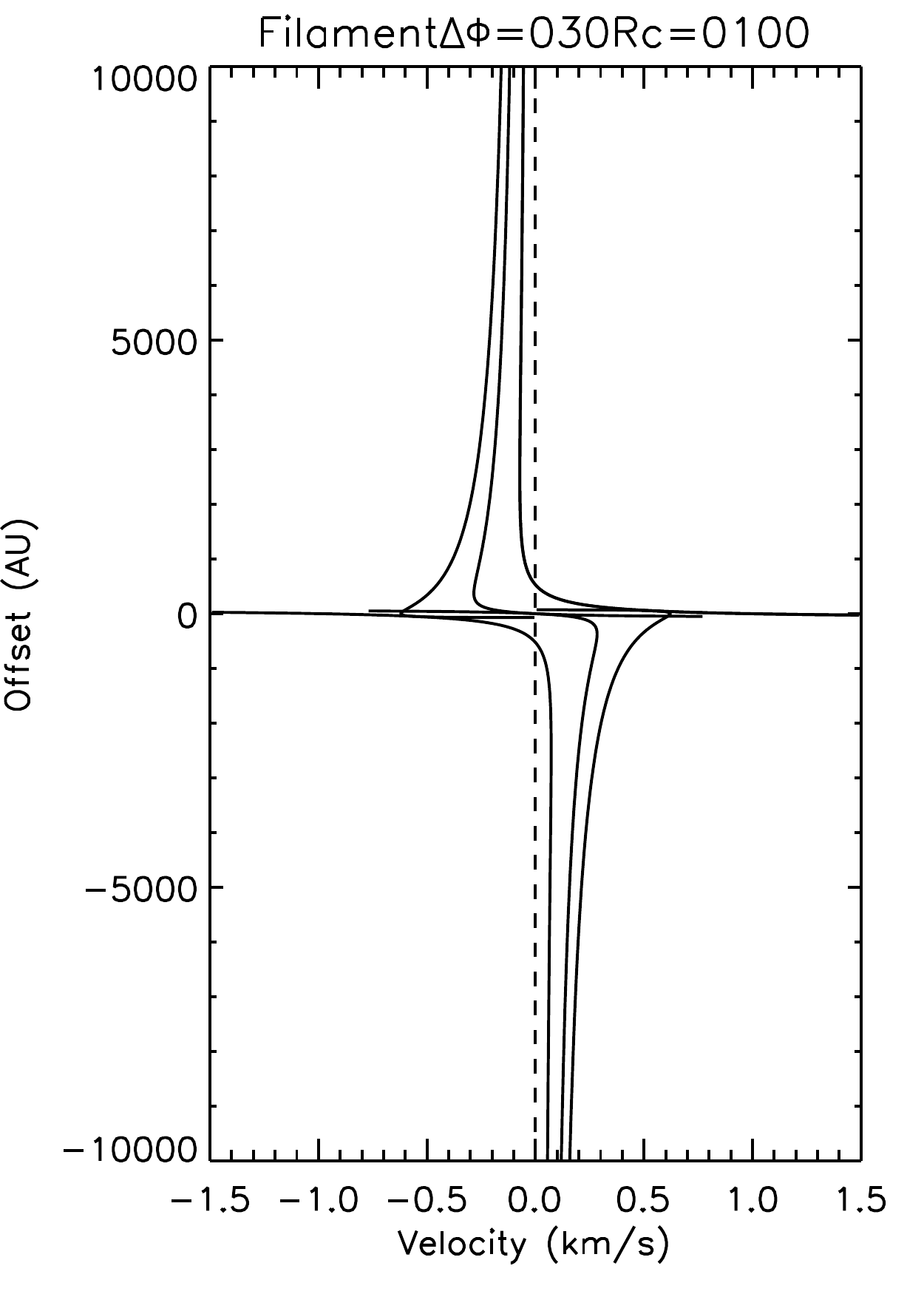}
\end{center}
\caption{Plots of infall streamlines for a filamentary envelope with $R_C$=100 AU. From top to bottom,
filamentary envelopes are shown with projection angles of $\Phi_0$= 0, -30, and 30\degr.
The \textit{left panel}
shows the streamlines out to 10000 AU while the \textit{middle panel} zooms in on the inner 1000 AU. 
The circle at the center represents the edge of the circumstellar disk forming at $R_C$. The \textit{right}
panel show the PV structure of the filamentary streamlines. The filamentary 
envelopes at a particular rotation angle can be thought of as showing a selected portion of the axisymmetric PV diagram.
}
\label{filament-streampv}
\end{figure}
\clearpage

\begin{figure}
\begin{center}
\includegraphics[scale=0.7,angle=-90]{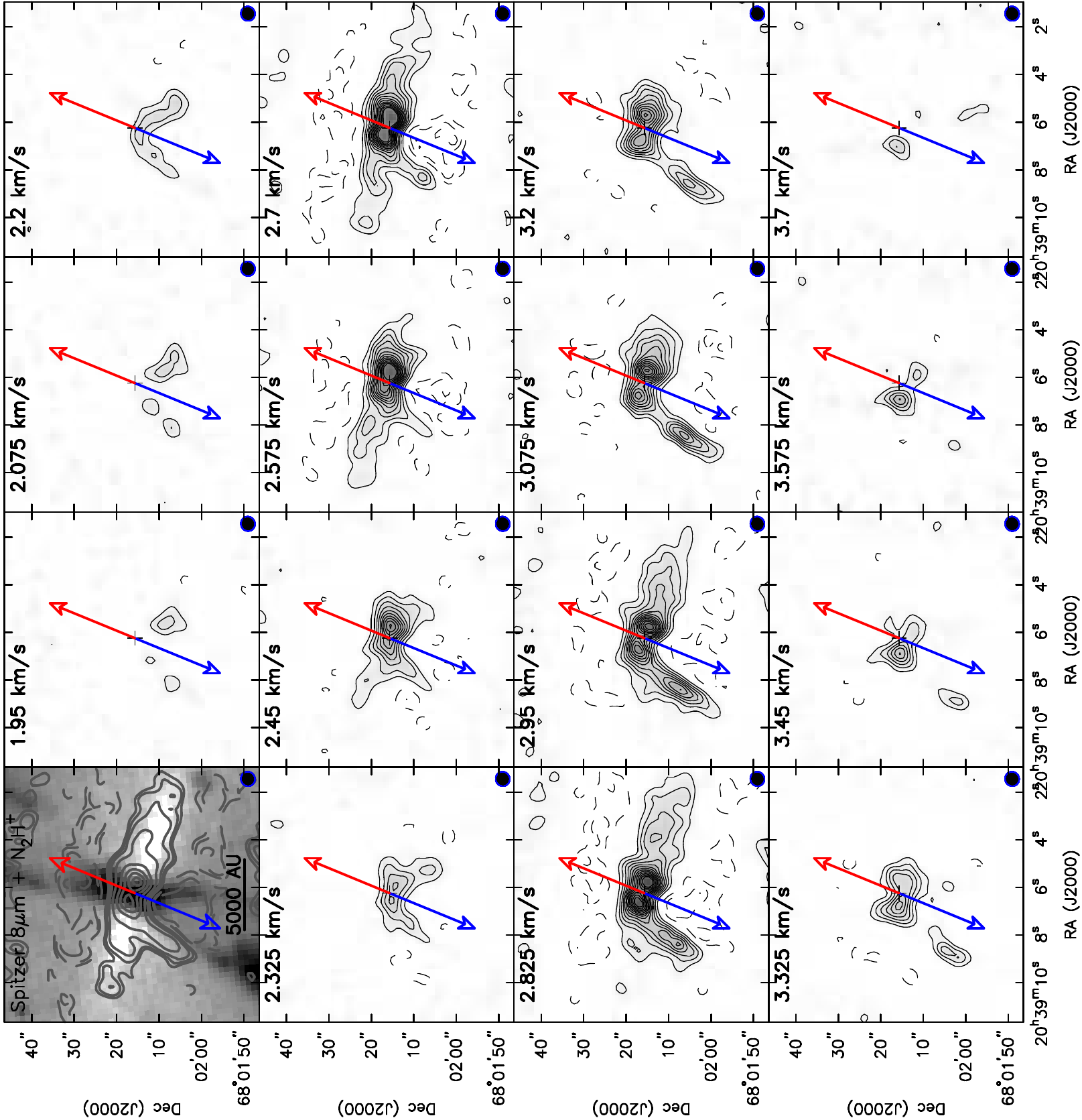}
\end{center}
\caption{L1157-- Channel maps of the isolated \nthp\ ($JF_1F = 101\rightarrow012$) line; the blue and red arrows
mark the outflow direction and the cross marks the location of the protostar. The higher velocity
data in the top and bottom rows mainly trace the likely outflow entrained material, while the middle two rows
primarily reflect emission due to envelope kinematics. Note the velocity gradient of the large-scale material
from 2.45 \kms\ to 3.075 \kms. The large-scale emission on the east and west sides of the envelope
has a rather constant velocity. The extension
of \nthp\ emission to the south traces an outflow cavity wall whose kinematics also reflect some outflow entrainment.
Contours in the channel maps start at $\pm$3$\sigma$ increase by $\pm$3$\sigma$ intervals, where $\sigma$=0.08 K. 
 The contours in the integrated
intensity map are $\pm$6, 9, 15, 30, 60$\sigma$, ..., where
$\sigma$=0.02 K \kms.} 
\label{L1157-chanmaps}
\end{figure}
\clearpage

\begin{figure}
\begin{center}
\includegraphics[scale=0.75,angle=-90]{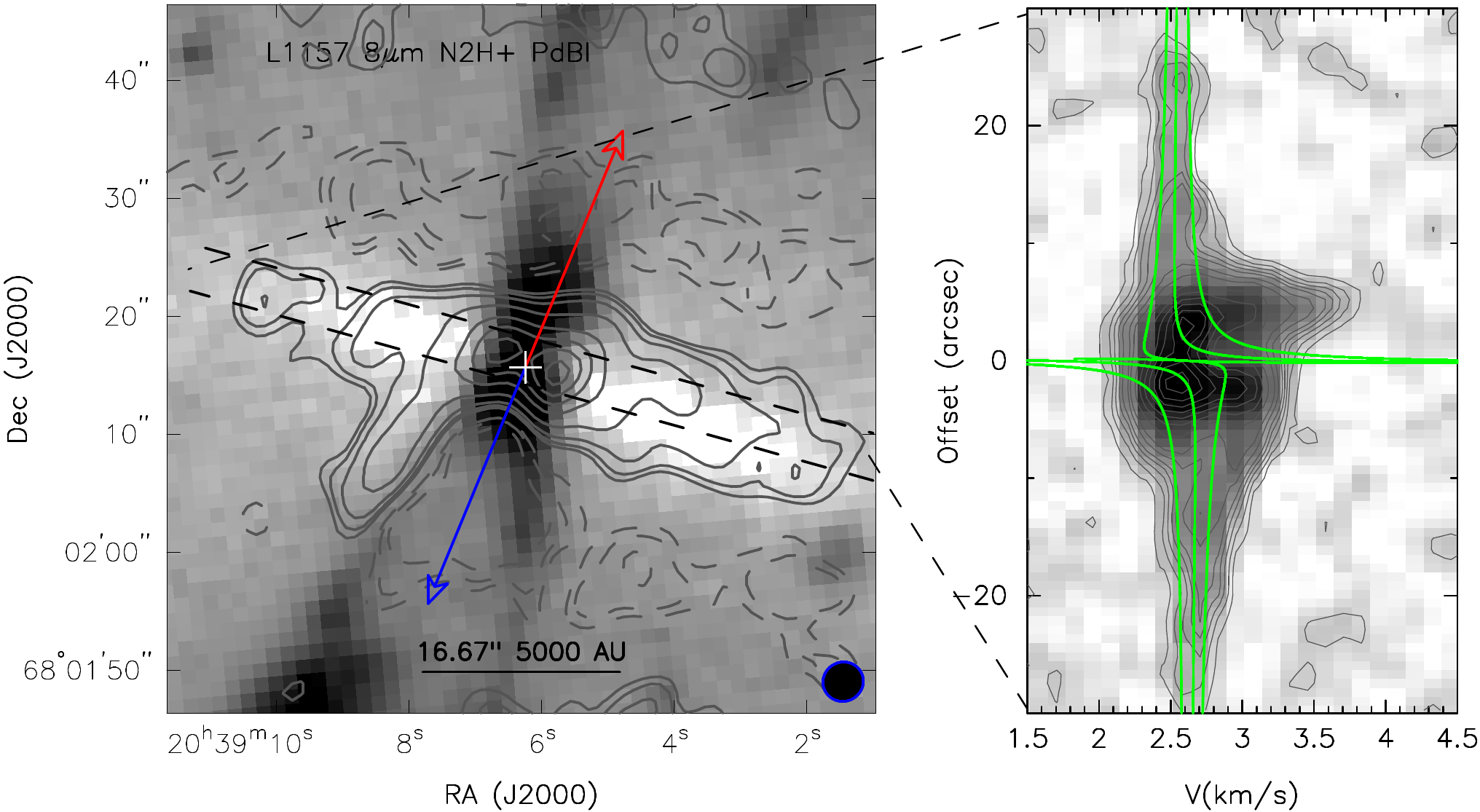}
\end{center}
\caption{L1157-- The \textit{left panel} shows the IRAC 8\mum\ image with PdBI \nthp\ integrated intensity contours overlaid
of the isolated \nthp\ line ($JF_1F = 101\rightarrow012$).
The dashed lines mark the regions where the position-velocity cut was taken and point to respective ends of the PV plot in
the \textit{right panel}. The position of the protostar/continuum source is marked with a white cross.
The position-velocity cut is shown for the isolated \nthp\ line,
showing that the two lumps of \nthp\ emission are directly associated with regions of increased
linewidth, likely due to outflow effects.  Outside of this region, the linewidth
becomes fairly narrow in the outer envelope with constant velocity.
The green lines overlaid are an filamentary
collapse model with $R_C$=100 AU and a projection of $\Phi_0$ = 15\degr.
The contours in the integrated
intensity map are $\pm$6, 9, 15, 30, 60$\sigma$, ..., where
$\sigma$=0.02 K \kms; the contours in the PV diagram are $\pm$3$\sigma$ and in increments of 3$\sigma$ where $\sigma$=0.17 K. } 
\label{L1157-pv-n2hp}
\end{figure}
\clearpage

\begin{figure}
\begin{center}
\includegraphics[scale=0.65,angle=-90]{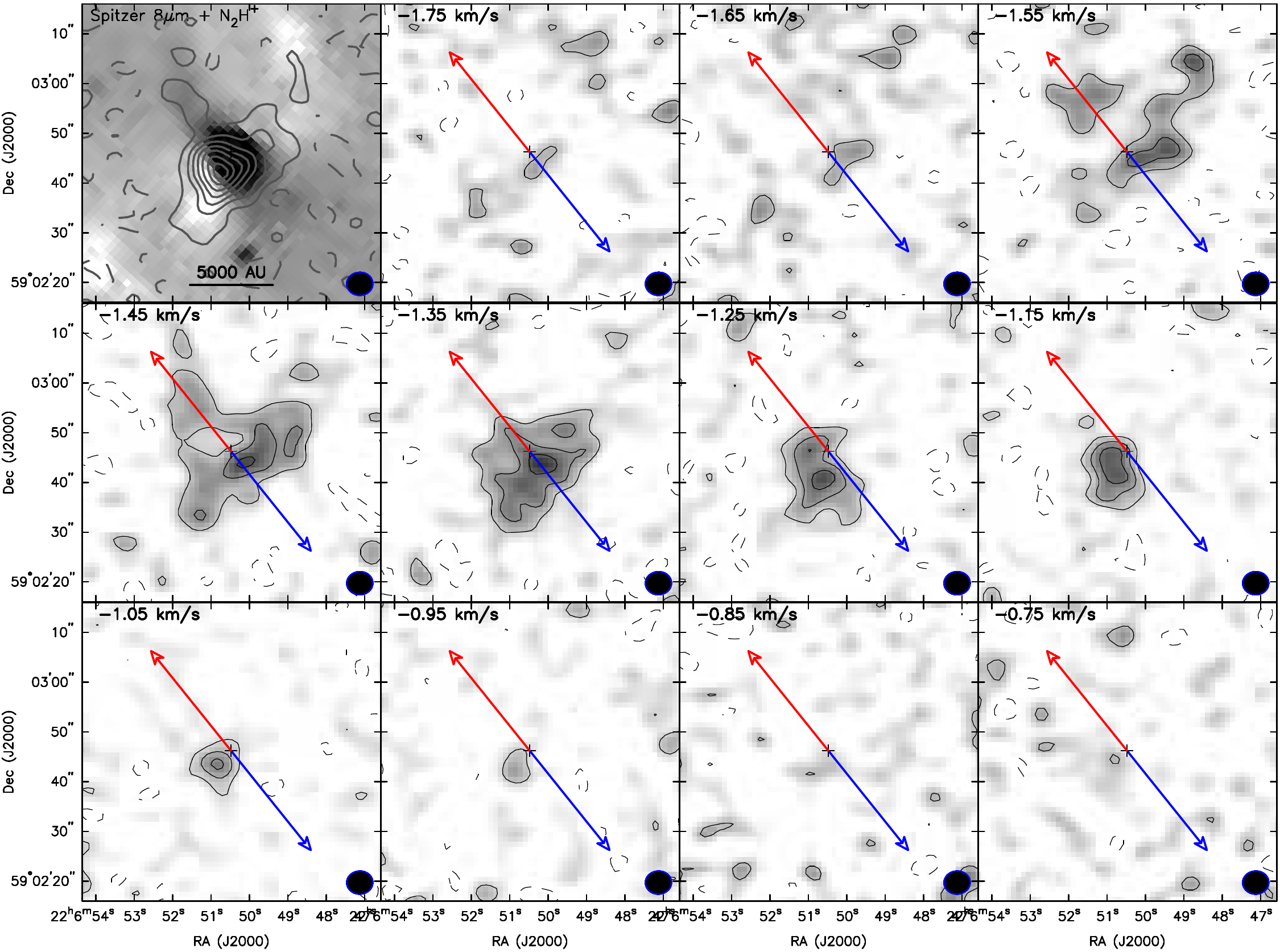}
\end{center}
\caption{L1165-- Channel maps of the isolated \nthp\ ($JF_1F = 122\rightarrow011$) line; the blue and red arrows
mark the outflow direction and the cross marks the location of the protostar.
 There is a rather well-behaved global velocity structure
in L1165 from blue-shifted northwest of the protostar and red-shifted to the southeast. There does not 
appear to be significant outflow effects on the kinematic structure. There is also higher
velocity red-shifted emission apparent on small-scales near the protostar, but very little blue-shifted emission
with higher velocities. 
The contours in the integrated intensity map are $\pm$2.5, 3, 6$\sigma$, ..., where
$\sigma$=0.207 K \kms; the contours in the channel maps are $\pm$3, 6$\sigma$, ..., where
$\sigma$=0.4 K.}
\label{L1165-chanmaps}
\end{figure}

\begin{figure}
\begin{center}
\includegraphics[scale=0.65,angle=-90]{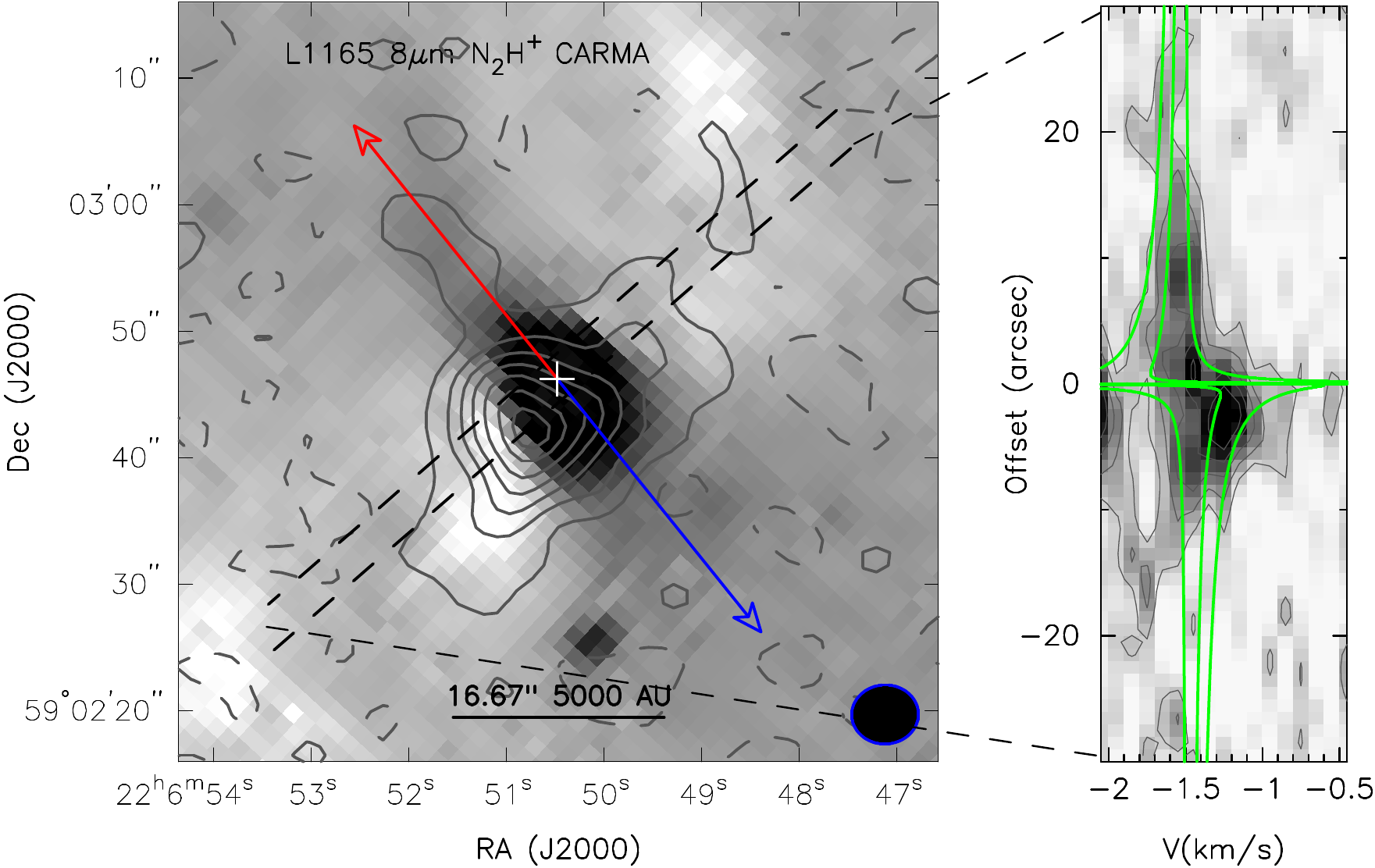}
\end{center}
\caption{L1165-- The \textit{left panel} shows the IRAC 8\mum\ image with CARMA \nthp\ integrated intensity contours overlaid.
The dashed lines mark the regions where the position-velocity cut was taken and point to respective ends of the PV plot in
the \textit{right panel} showing the \nthp\ ($JF_1F = 122\rightarrow011$) transition. The position of
the protostar/continuum source is marked with a white cross.
Northwest of the protostar the lines tend to be more narrow than on the southeast side of the protostar where
the \nthp\ emission is both broad and red-shifted.
The green lines overlaid are an filamentary
collapse model with $R_C$=10 AU and a projection of $\Phi_0$ = 15\degr.
The contours in the integrated intensity map are $\pm$2.5, 3, 6$\sigma$, ..., where
$\sigma$=0.207 K \kms; the contours in the PV diagram are $\pm$3$\sigma$, increasing in increments of 3$\sigma$ where $\sigma$=0.225 K.}
\label{L1165-pv-n2hp}
\end{figure}

\begin{figure}
\begin{center}
\includegraphics[scale=0.75,angle=-90]{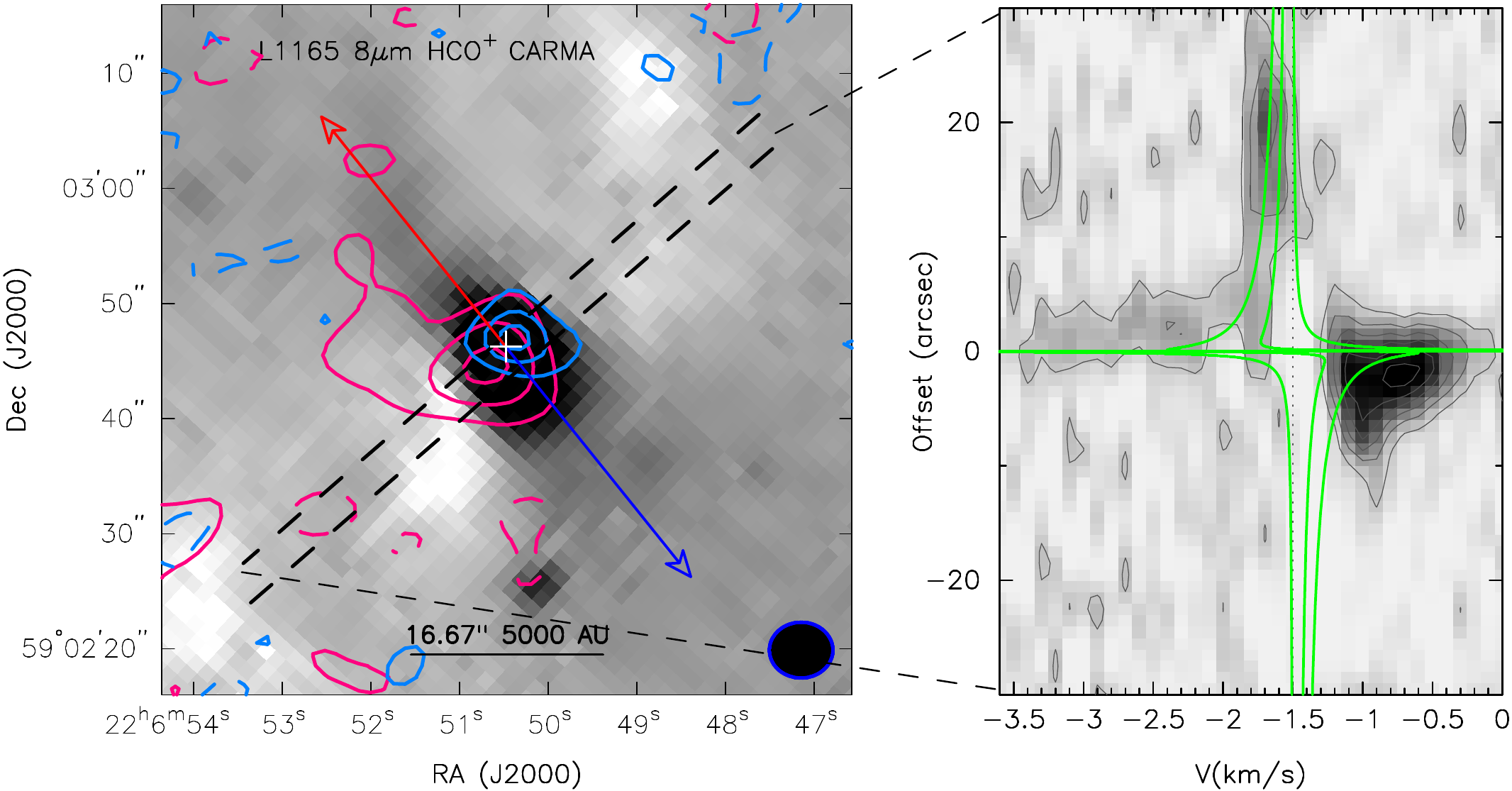}
\end{center}
\caption{L1165-- The \textit{left panel} shows the IRAC 8\mum\ image with CARMA HCO$^+$ (J=1$\rightarrow0$)
blue and red-shifted emission, summed over -3.5 to -2 \kms\ and -1.0 to 0 \kms, 
plotted as blue and red contours respectively. The contours levels are 3, 6, and 8.25$\sigma$ ($\sigma$=0.175 K)
for the blue-shifted emission and 3, 9, and 18$\sigma$ ($\sigma$=0.212 K) for the red-shifted emission.
 The blue and red-shifted emission
from HCO$^+$ is located symmetrically about
the protostar, normal to the outflow. The dashed lines mark the regions where
the position-velocity cut was taken and point to respective ends of the PV plot in
the \textit{right panel}. The position of the protostar/continuum source is marked
with a white cross. The position-velocity cut shows that the blue and red-shifted
emission traces higher velocity material and there is a slight gradient of material
going to higher velocity closer to the continuum source. The line center velocity of -1.5 \kms\
is plotted as the dotted black line. 
The green lines overlaid are an filamentary
collapse model with $R_C$=10 AU and a projection of $\Phi_0$ = 15\degr.
The PV plot contours start at 3$\sigma$
and increase in 3$\sigma$ intervals ($\sigma$=0.2).}
\label{L1165-pv-hco}
\end{figure}
\clearpage

\begin{figure}
\begin{center}
\includegraphics[scale=0.65,angle=-90]{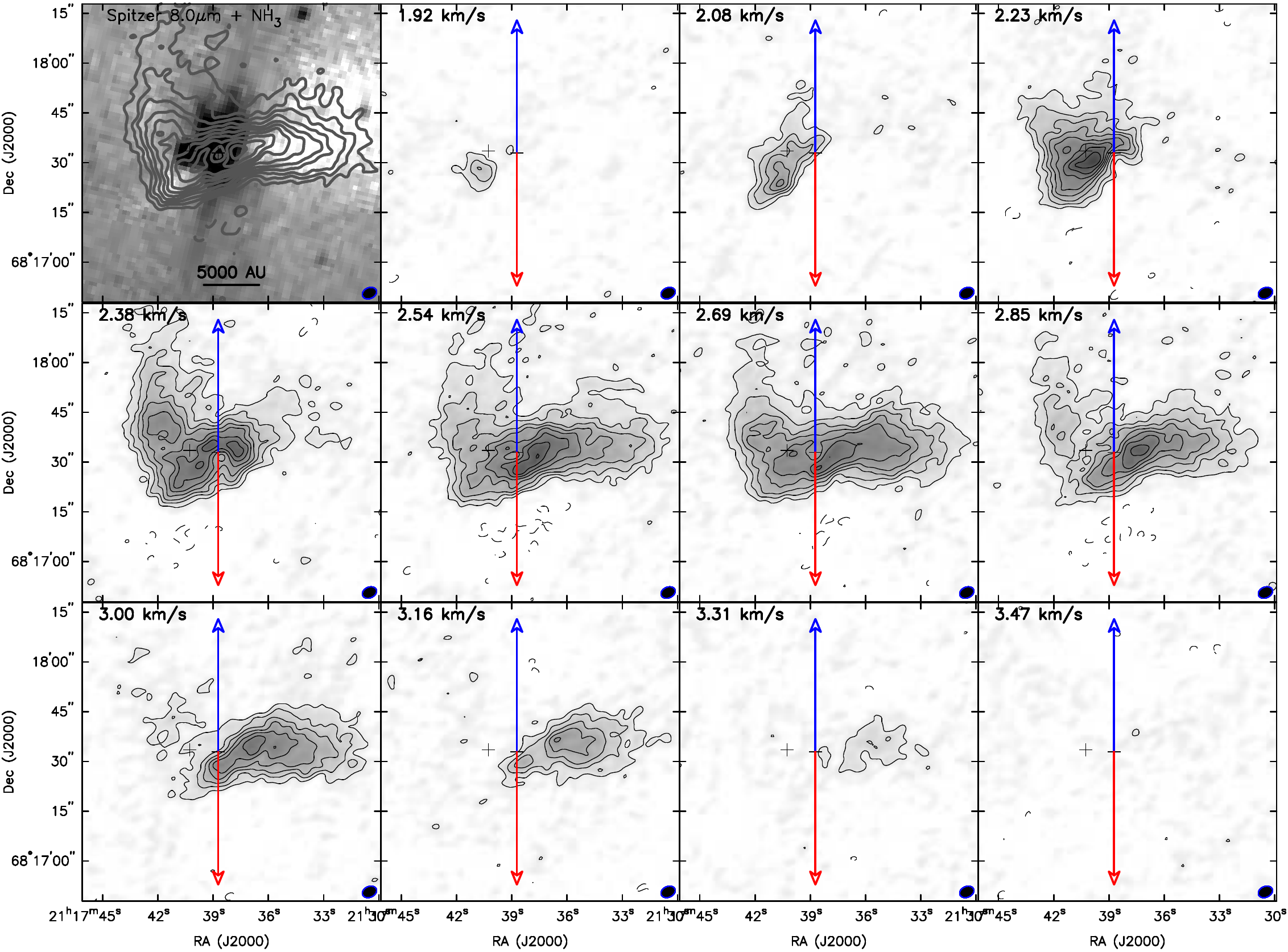}
\end{center}
\caption{CB230-- \nht\ integrated intensity map of the main \nht\ (1,1) lines (\textit{upper left}) and 
channel maps of the two main \nht\ (1,1) lines are shown in the rest of the panels. The protostars are marked 
with crosses and the outflow direction is drawn from
the main protostar. The emission between 2.54 and 2.85 \kms\ consists of blended emission
from both lines, while the remaining emission is mostly free of blending. The channel maps
show a clear velocity gradient across the source, appearing normal to the outflow. Notice how the
emission tends to avoid the location of the main protostar, owing to destruction of \nht\ near the protostar.
The contours in the integrated intensity map start at $\pm$3$\sigma$, increasing in increments of 3$\sigma$ and
$\sigma$=0.195 K \kms; the contours in the integrated intensity map are $\pm$3, 6$\sigma$, ..., where
$\sigma$=0.325 K.}
\label{CB230-chanmaps}
\end{figure}

\begin{figure}
\begin{center}
\includegraphics[scale=0.75,angle=-90]{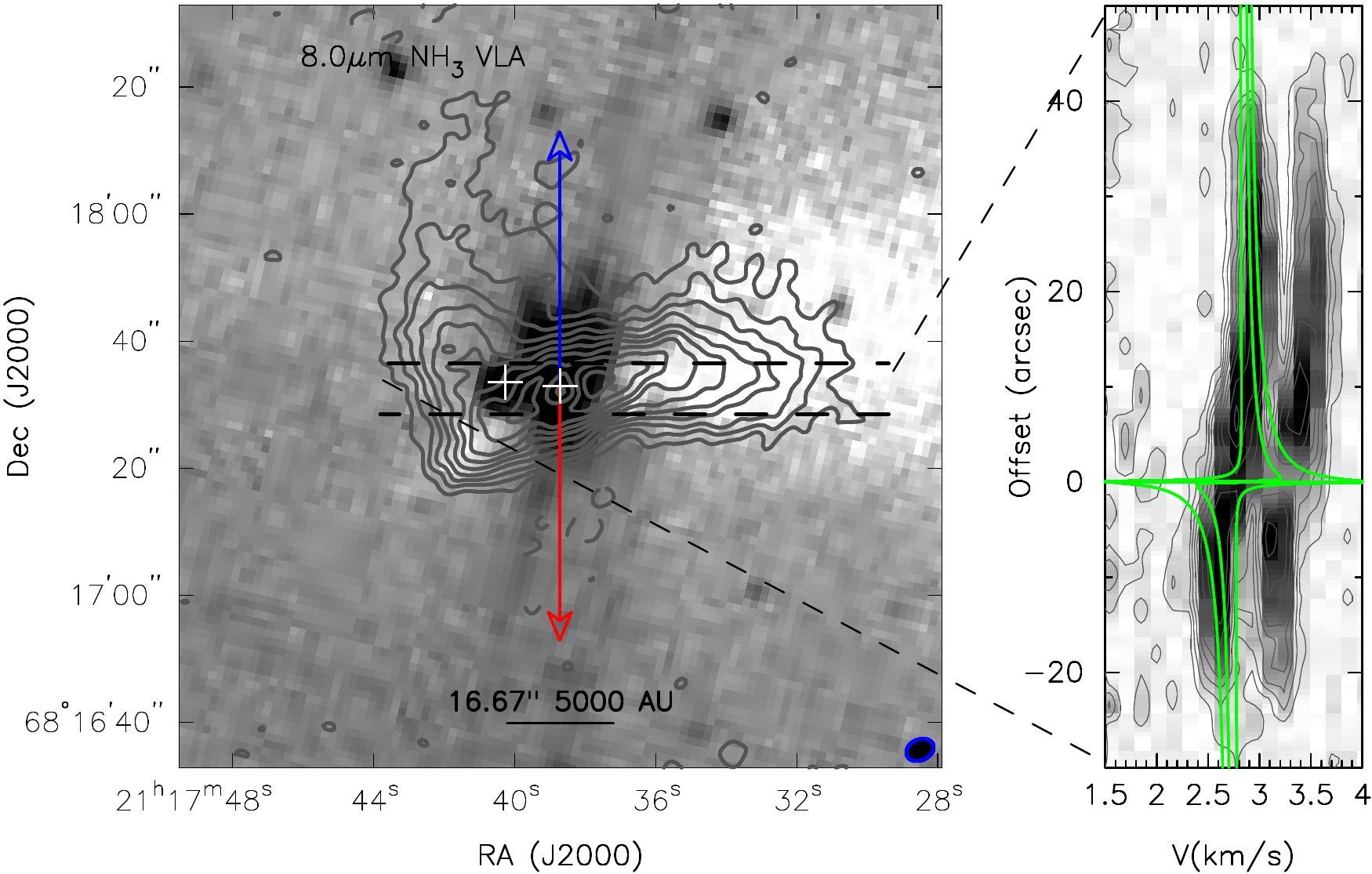}

\end{center}
\caption{CB230-- Same as Figure \ref{L1157-pv-n2hp} but with VLA \nht\ (1,1) observations
and the secondary source is also marked with a white cross. The PV
diagram in the \textit{right panel} appears different from those constructed from \nthp\ because
the \nht\ emission consists of two blended hyperfine lines; the \nht\ satellite lines
with greater separation are shown in the PV plot rather than the more blended main lines. The line-center velocity
from the \nht\ emission traces an abrupt velocity shift
from red to blue-shifted emission. The transition region starts at the protostar and
does not finish until $+$10\arcsec. The \nht\ linewidth does not show much detail other than
having is peak coincident with the highest intensity \nht\ emission.
The green lines overlaid are an filamentary
collapse model with $R_C$=10 AU and a projection of $\Phi_0$ = 20\degr.
The contours in the integrated intensity map start at $\pm$3$\sigma$, increasing in increments of 3$\sigma$ and
$\sigma$=0.195 K \kms; the contours in the PV diagram start at $\pm$3$\sigma$, increasing in increments of 3$\sigma$ and $\sigma$=0.11 K.
}
\label{CB230-pv-nh3}
\end{figure}
\clearpage

\begin{figure}
\begin{center}
\includegraphics[scale=0.75,angle=-90]{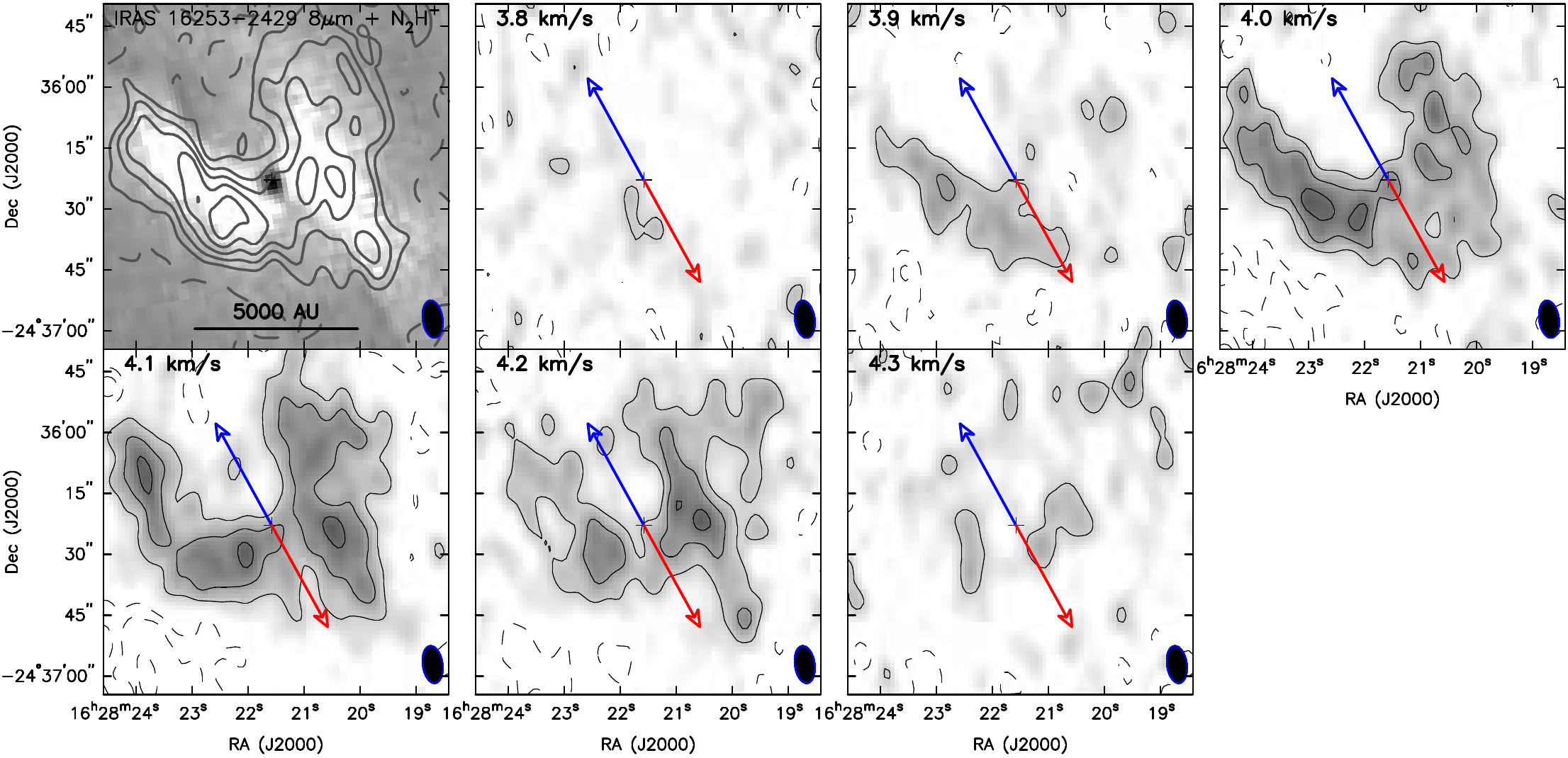}

\end{center}
\caption{IRAS 16253-2429-- Channel maps of the isolated \nthp\ ($JF_1F = 122\rightarrow011$) line; the blue and red arrows
mark the outflow direction and the cross marks the location of the protostar.
 IRAS 16253-2429 is the envelope that is most nearly round. The velocity
gradient across the envelope is very small, but is clearing seen in the channel maps with 
emission on the eastern side of the envelope coming into view at blueshifted velocities and the 
western side with redshifted velocities. The green lines overlaid are an axisymmetric
collapse model with $R_C$=100 AU and $M_C$ = 0.1 $M_{\sun}$.
Contours in the channel maps start at $\pm$3$\sigma$ increase 
by $\pm$3$\sigma$ intervals, where $\sigma$=0.08 K. 
The contours in the integrated intensity map are $\pm$2.5, 3, 6$\sigma$, ..., where
$\sigma$=0.207 K \kms.}
\label{IRAS16253-chanmaps}
\end{figure}
\clearpage

\begin{figure}
\begin{center}
\includegraphics[scale=0.75,angle=-90]{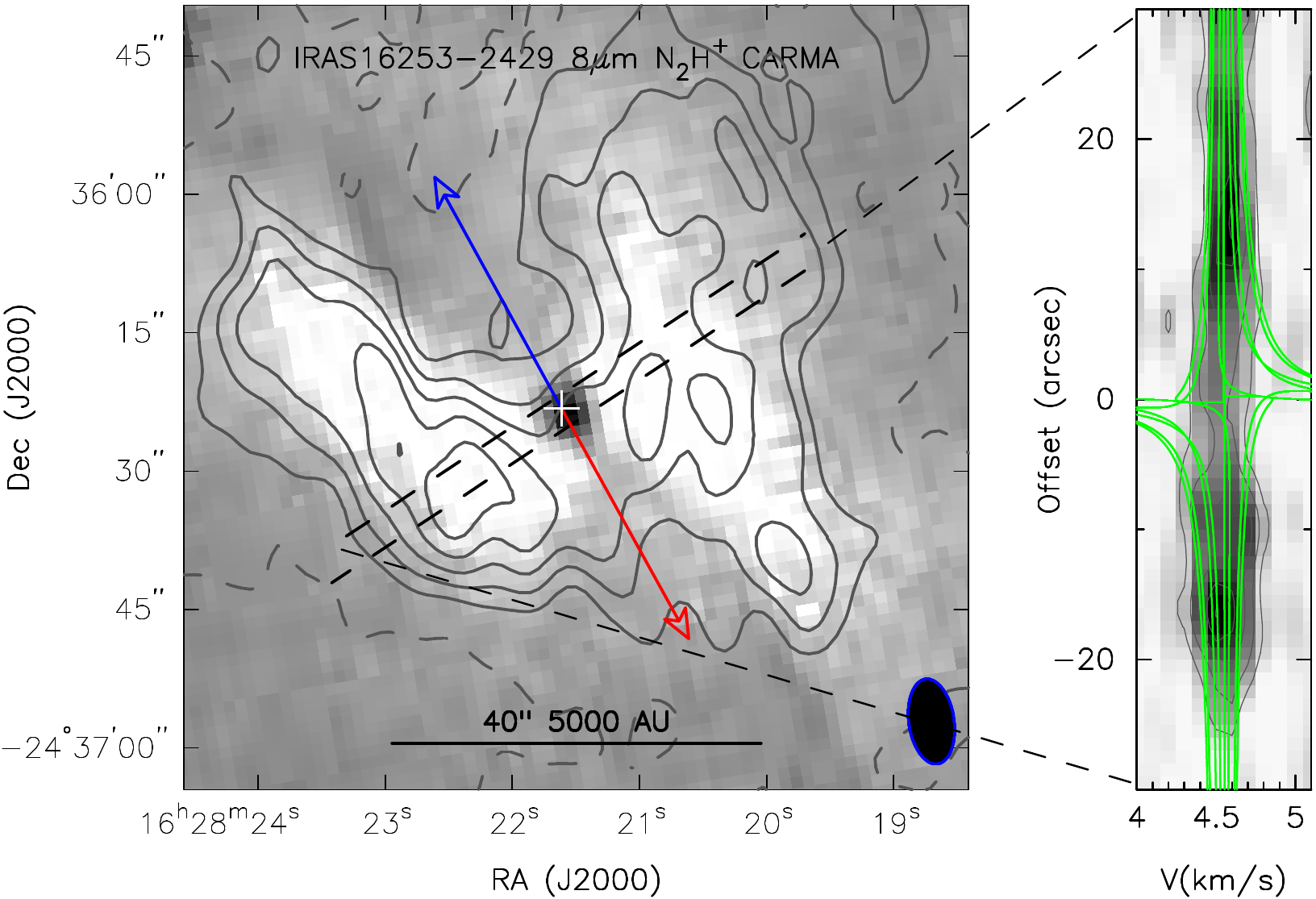}

\end{center}
\caption{IRAS 16253-2429-- Same as Figure \ref{L1157-pv-n2hp} but with CARMA \nthp\ observations 
of the ($JF_1F = 122\rightarrow011$) transition.
The \nthp\ emission matches up quite well with the 8\mum\ extinction. Furthermore, the \nthp\ appears peaked around
the protostar and not coincident with it. This is shown as the decreased emission in the PV plot
in the right panel. The velocity gradient in the PV plot is minute but is present.
The green lines overlaid are an filamentary
collapse model with $R_C$=10 AU and a projection of $\Phi_0$ = 15\degr.
The contours in the integrated intensity map start at $\pm$3$\sigma$, increasing in increments of 3$\sigma$ and
$\sigma$=0.22 K \kms; the contours in the PV diagram start at $\pm$3$\sigma$, increasing in increments of 3$\sigma$ and $\sigma$=0.25 K.}
\label{IRAS16253-pv-n2hp}
\end{figure}
\clearpage

\begin{figure}
\begin{center}
\includegraphics[scale=0.75,angle=-90]{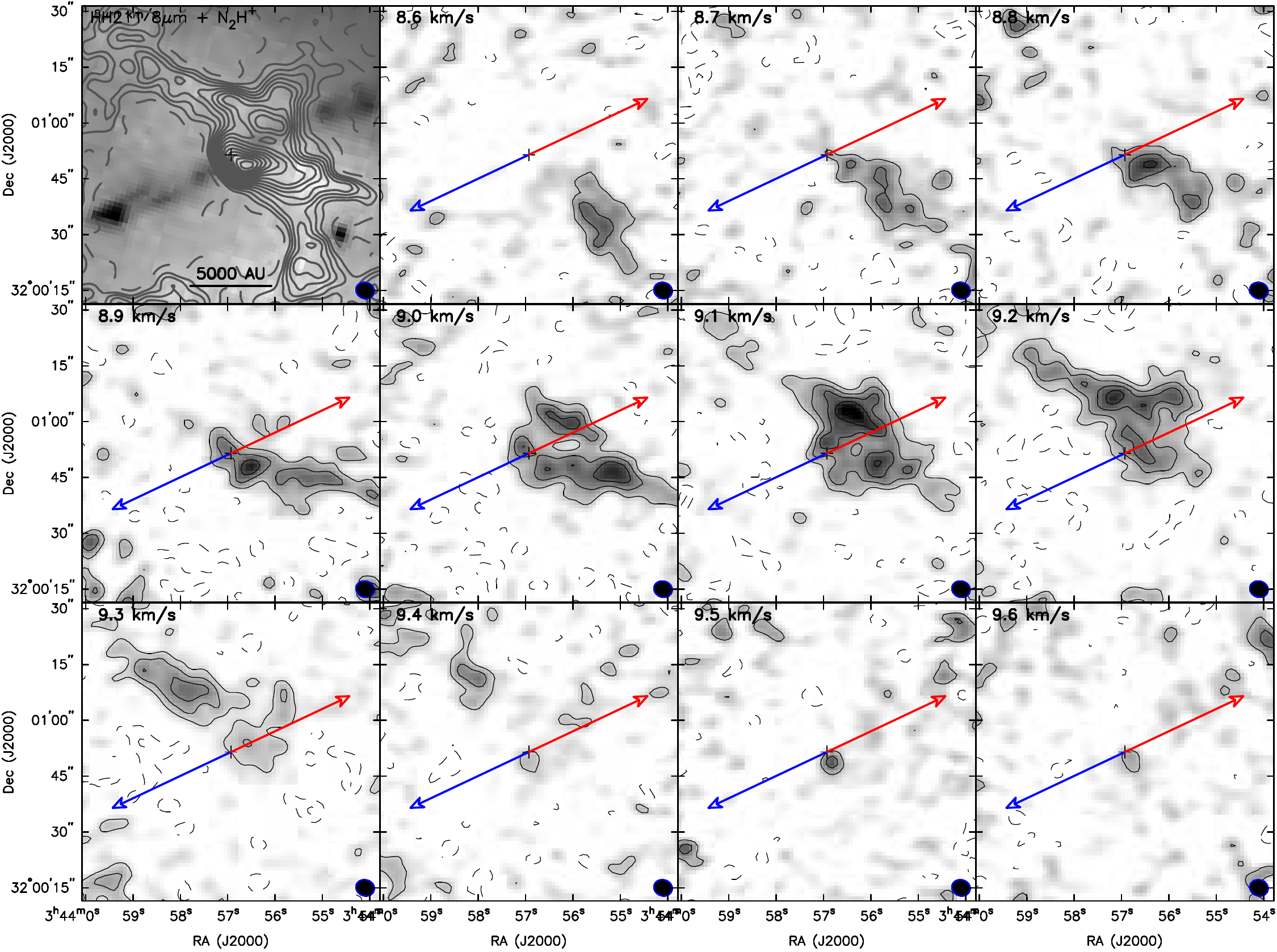}
\end{center}
\caption{HH211-- Channel maps of the isolated \nthp\ ($JF_1F = 101\rightarrow012$) line; the blue and red arrows
mark the outflow direction and the cross marks the location of the protostar.
The \nthp\ emission in the different velocity channels clearly trace a filamentary envelope with
a velocity gradient across the protostar that appears fairly linear.
Contours in the channel maps start at $\pm$3$\sigma$ increase by $\pm$3$\sigma$ intervals, where $\sigma$=0.4 K. 
The contours in the integrated intensity map are $\pm$10, 20$\sigma$, ..., where
$\sigma$=0.149 K \kms.} 
\label{HH211-chanmaps}
\end{figure}
\clearpage

\begin{figure}
\begin{center}
\includegraphics[scale=0.75,angle=-90]{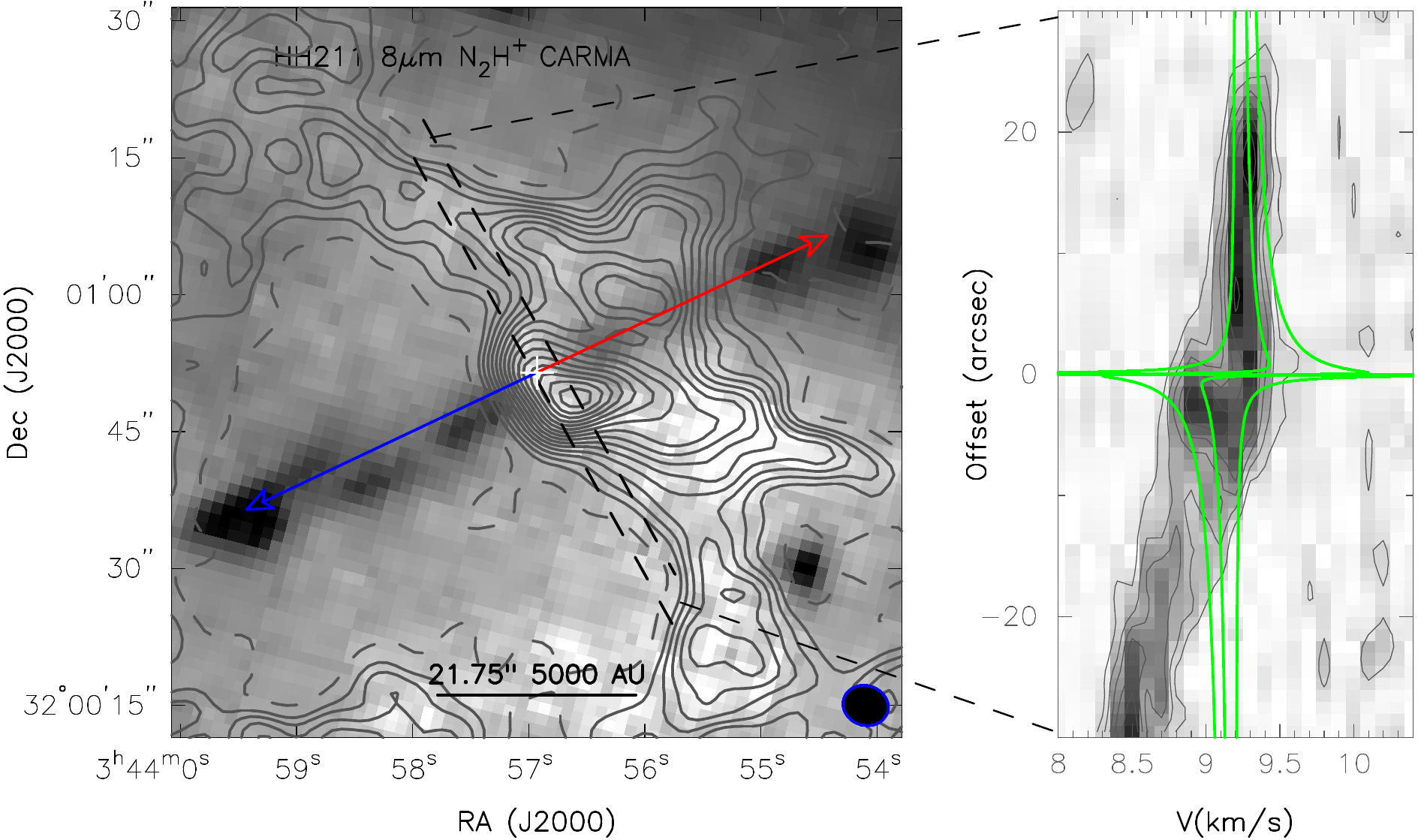}
\end{center}
\caption{HH211-- Same as Figure \ref{L1157-pv-n2hp} but with CARMA \nthp\ observations.
 The PV plot in the \textit{right panel} is constructed from the \nthp\ ($JF_1F = 101\rightarrow012$) transition.
Notice how the slope of emission in the PV diagram changes from one side of the protostar to the other.
The green lines overlaid are an filamentary
collapse model with $R_C$=10 AU and a projection of $\Phi_0$ = 15\degr\ that are meant
to represent the kinematics on the northeast side of the envelope.
The contours in the integrated intensity map start at $\pm$10$\sigma$, increasing in increments of 10$\sigma$ and
$\sigma$=0.149 K \kms; the contours in the PV diagram start at $\pm$3$\sigma$, increasing in increments of 3$\sigma$ and $\sigma$=0.2 K.
}
\label{HH211-pv-n2hp}
\end{figure}
\clearpage

\begin{figure}
\begin{center}
\includegraphics[scale=0.4]{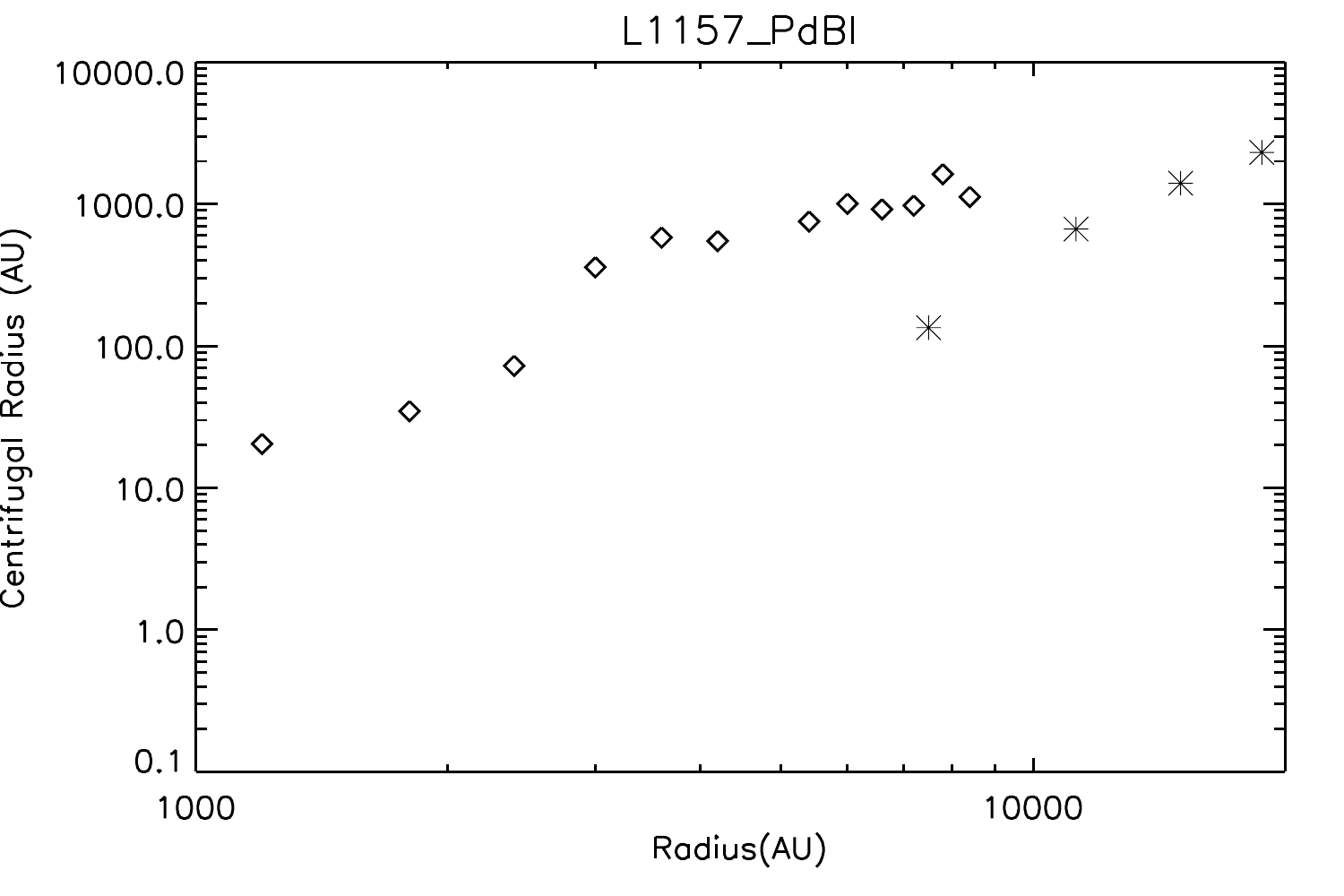}
\includegraphics[scale=0.4]{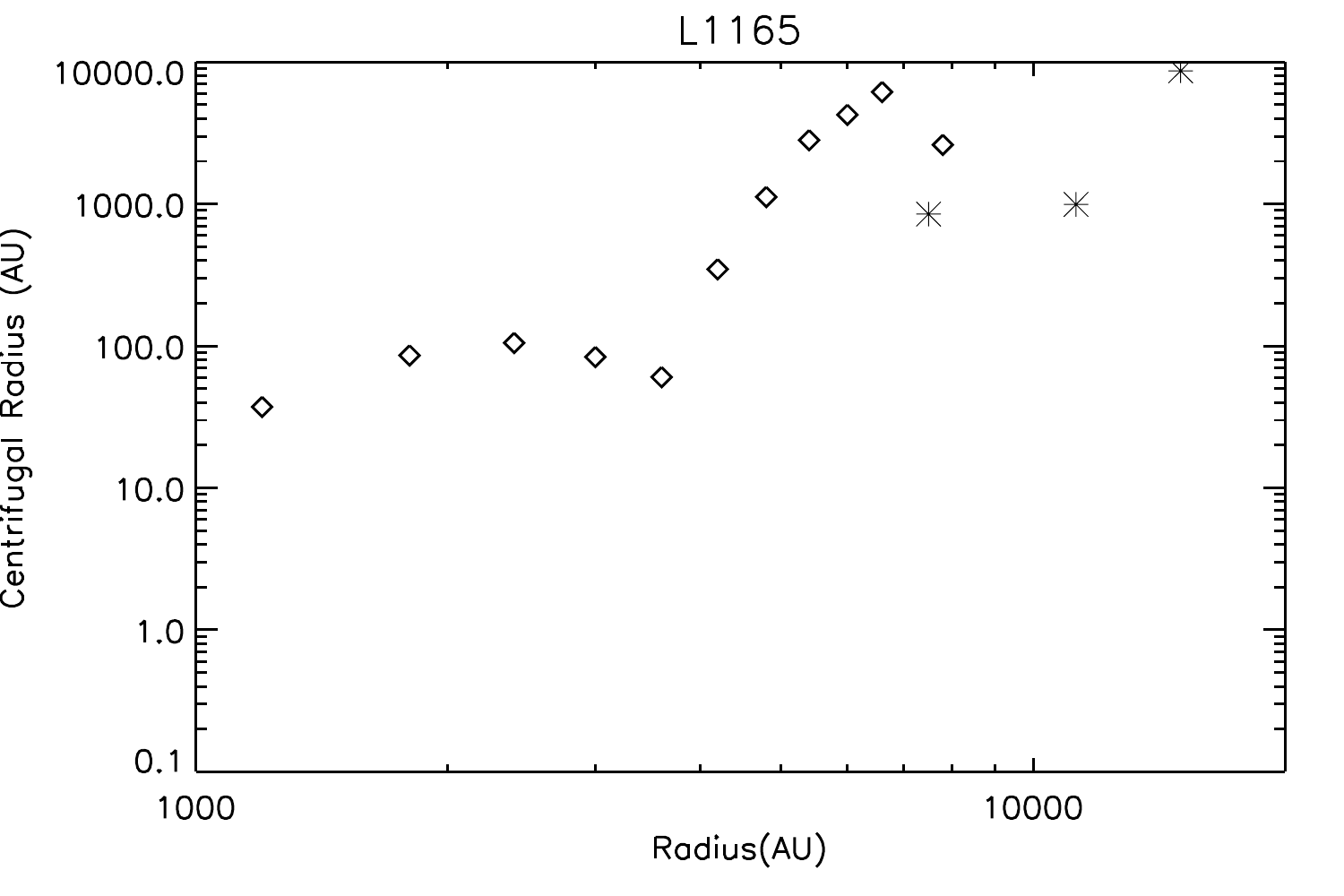}
\includegraphics[scale=0.4]{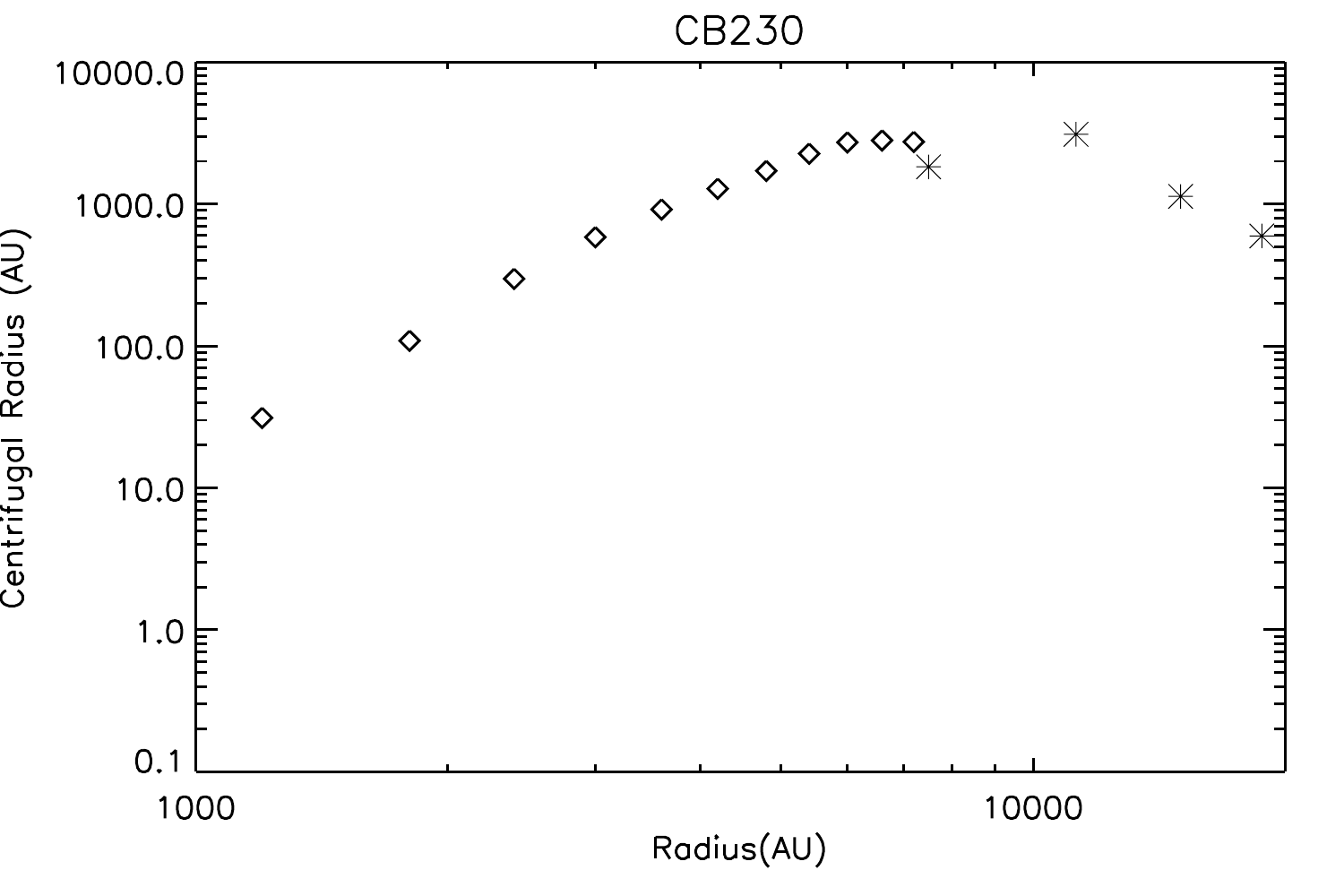}
\includegraphics[scale=0.4]{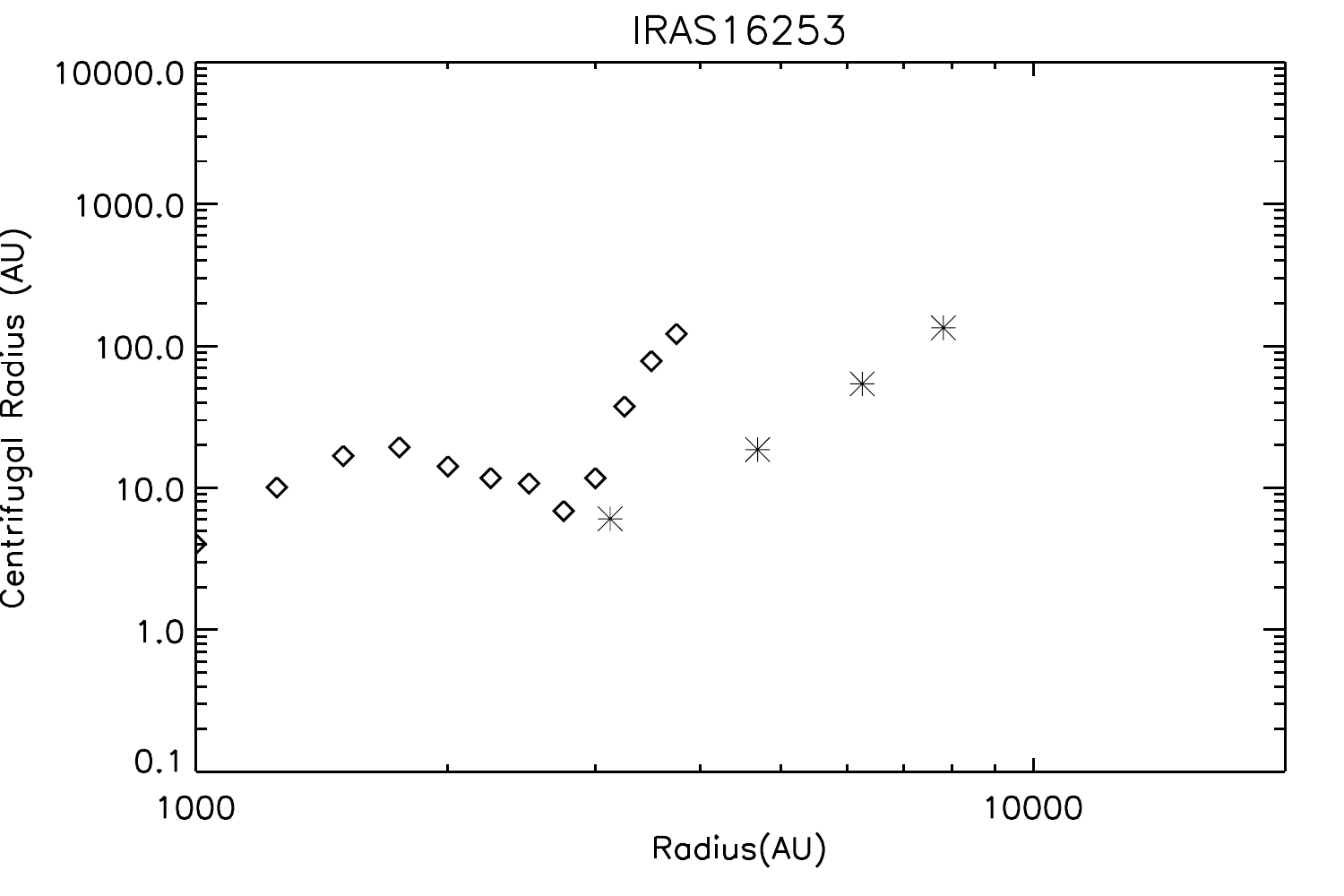}
\includegraphics[scale=0.4]{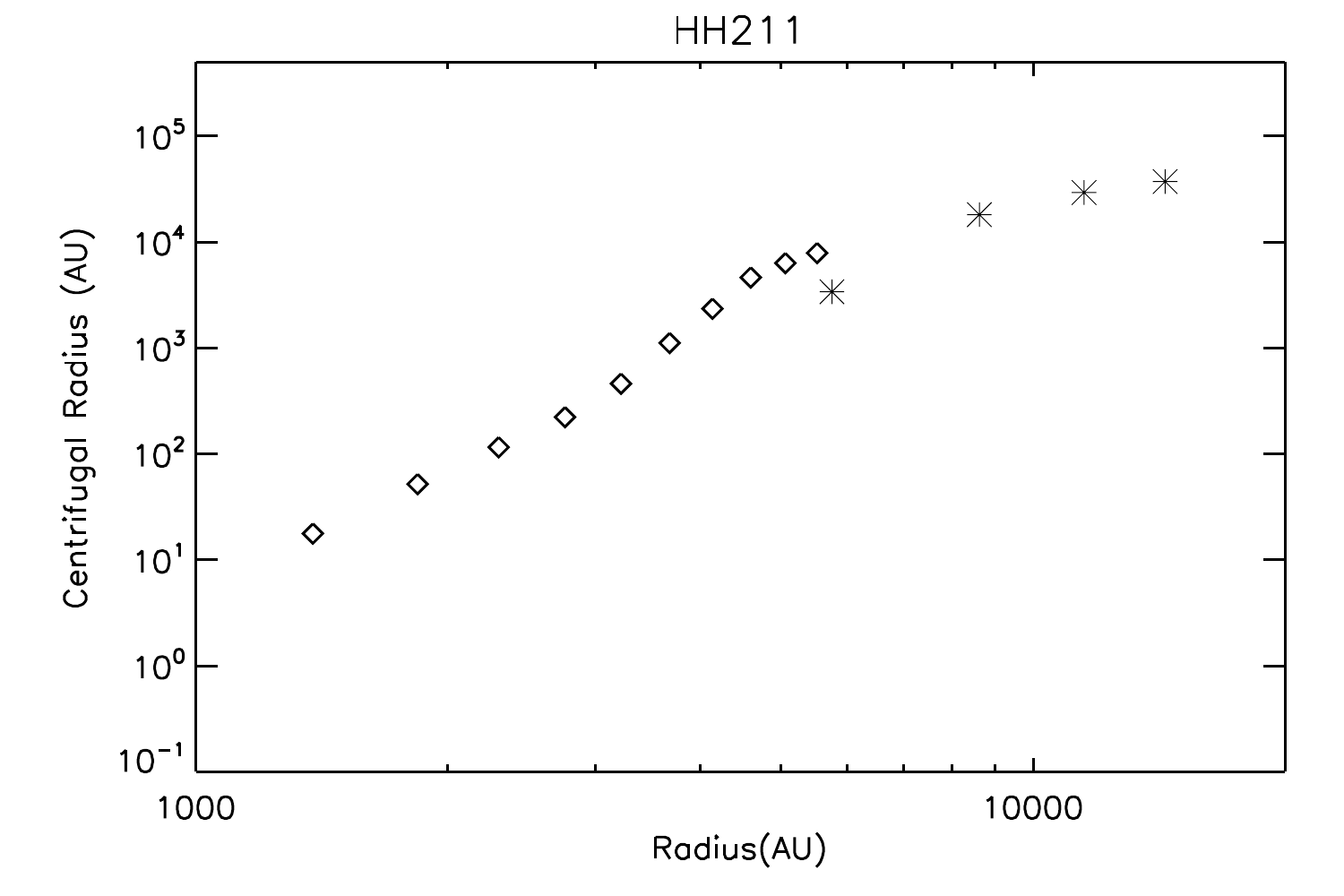}
\end{center}
\caption{Plot of centrifugal radius ($R_C$) versus radius for L1157, L1165, CB230, IRAS 16253-2429, and HH211 assuming the 
observed velocities reflect rotation. $R_C$ is calculated from the 
observed line-center velocities from Paper II and converted to a centrifugal radius using the relation
$R_C$ = $R^2(V-V_{lsr})^2/GM$ (Equation 4), where $M$ $=$ 0.5 $M_{\sun}$ is assumed.}
\label{RCR}
\end{figure}
\end{small}
\clearpage

\input{tab1}
\input{tab2}

\end{document}

%% file: tab1.tex
\begin{deluxetable}{llllllllllll}
\rotate
\tablewidth{0pt}
\tabletypesize{\scriptsize}
\tablecaption{Source Properties}
\tablehead{
  \colhead{Source}  & \colhead{RA} & \colhead{Dec}      & \colhead{Distance}    &  \colhead{Mass$_{8\mu m}$} & \colhead{Mass$_{submm}$\tablenotemark{*}}   & \colhead{L$_{bol}$}     & \colhead{T$_{bol}$}  &\colhead{Morphological} & \colhead{Outflow PA} & \colhead{References}\\
                    & \colhead{(J2000)} &  \colhead{(J2000)}     & \colhead{(pc)}          &  \colhead{($M_{\sun}$)}     & \colhead{($M_{\sun}$)}                      & \colhead{($L_{\sun}$)}  & \colhead{(K)}        &\colhead{Classification} & \colhead{\degr} & \colhead{(Dist., M$_{ref}$, L$_{bol}$, T$_{bol}$)}  \\
                    &                   &     &                                           &  \colhead{(r$<$0.05pc)}     &  & & & &
}
\startdata
L1157           & 20:39:06.25 & +68:02:15.9 & 300 & 0.86  & 2.2        & 3.0  & 29     & Flattened  & 150 & 4, 3, 2, 9\\
L1165           & 22:06:50.46 & +59:02:45.9  & 300 & 1.1  & 0.32        & 13.9 & 46     & Irregular & 225 & 4,  6, 6, 1\\
CB230           & 21:17:38.56 & +68:17:33.3 & 300  & 1.1  & 1.1        & 7.2   & 69     & One-sided & 0 &   4, 8, 2, 9\\
HH211           & 03:43:56.78 & +32:00:49.8 & 230  & 1.1 &   1.5        & 3.02  & 24     & Irregular & 116 & 12, 2, 2, 2, 5  \\
IRAS 16253-2429 & 16:28:21.42 & -24:36:22.1 & 125 & 0.8  & 0.98         & 0.25 & 35     & Spheroidal & 20 & 7, 10, 2, 2, 11\\
\enddata
\tablecomments{Properties of sources observed in the single-dish and/or interferometric sample. The 8\mum\ extinction
masses are taken within 0.05 pc of the protostar and note that some of the masses have been rescaled to account 
for a different distance estimate as compared to Chapter 4. Positions are reflect the coordinates of the 24\mum\ 
point source from \textit{Spitzer} data or the 3mm continuum continuum position for protostars observed with CARMA.
References: (1) \citet{tobin2011}, (2) \citet{enoch2009}, 
 (3) \citet{shirley2000}, (4) \citet{kirk2009}, (5) \citet{lee2009}, (6) \citet{visser2002}, (7) \citet{loinard2008},
(8)  \citet{kauffmann2008}, (9) \citet{froebrich2005}, (10) \citet{young2006}, (11) \citet{arce2006},(12) \citet{hirota2011}.
}
\tablenotetext{*}{Mass was computed with sub/millimeter bolometer data assuming an isothermal temperature.}

\end{deluxetable}

%% file: tab2.tex
\begin{deluxetable}{lllll}

\tablewidth{0pt}
\tabletypesize{\scriptsize}
\tablecaption{Model Parameters}
\tablehead{
  \colhead{Source}  & \colhead{Model} & \colhead{$M_{cent}$} & \colhead{$R_{C}$} & \colhead{Proj. Angle} \\
                    &                 & \colhead{$M_{\sun}$} & \colhead{(AU)}    & \colhead{($^{\circ}$)}\\
}
\startdata
L1157              &  Filament              & 0.5     & 100            &  15  \\
L1165              &  Filament              & 0.5     & 10             &  15  \\
CB230              &  Filament              & 0.5     & 10             &  20 \\
IRAS 16253-2429    &  Filament/Axisymmetric & 0.1     & 100            & 0 \\
HH211              &  Filament              & 0.5     & 10             &  15   \\

\enddata
\tablecomments{Parameters of models which describe the PV data. } 

\end{deluxetable}

%% file: ms.bbl
\begin{thebibliography}{62}
\expandafter\ifx\csname natexlab\endcsname\relax\def\natexlab#1{#1}\fi

\bibitem[{{Andrews} \& {Williams}(2007)}]{andrews2007}
{Andrews}, S.~M., \& {Williams}, J.~P. 2007, \apj, 659, 705

\bibitem[{{Andrews} {et~al.}(2009){Andrews}, {Wilner}, {Hughes}, {Qi}, \&
  {Dullemond}}]{andrews2009}
{Andrews}, S.~M., {Wilner}, D.~J., {Hughes}, A.~M., {Qi}, C., \& {Dullemond},
  C.~P. 2009, \apj, 700, 1502

\bibitem[{{Arce} \& {Sargent}(2006)}]{arce2006}
{Arce}, H.~G., \& {Sargent}, A.~I. 2006, \apj, 646, 1070

\bibitem[{{Arquilla} \& {Goldsmith}(1986)}]{arquilla1986}
{Arquilla}, R., \& {Goldsmith}, P.~F. 1986, \apj, 303, 356

\bibitem[{{Bacmann} {et~al.}(2000){Bacmann}, {Andr{\'e}}, {Puget}, {Abergel},
  {Bontemps}, \& {Ward-Thompson}}]{bacmann2000}
{Bacmann}, A., {Andr{\'e}}, P., {Puget}, J., {Abergel}, A., {Bontemps}, S., \&
  {Ward-Thompson}, D. 2000, \aap, 361, 555

\bibitem[{{Bate} {et~al.}(2003){Bate}, {Bonnell}, \& {Bromm}}]{bate2003}
{Bate}, M.~R., {Bonnell}, I.~A., \& {Bromm}, V. 2003, \mnras, 339, 577

\bibitem[{{Belloche} \& {Andr{\'e}}(2004)}]{belloche2004}
{Belloche}, A., \& {Andr{\'e}}, P. 2004, \aap, 419, L35

\bibitem[{{Belloche} {et~al.}(2002){Belloche}, {Andr{\'e}}, {Despois}, \&
  {Blinder}}]{belloche2002}
{Belloche}, A., {Andr{\'e}}, P., {Despois}, D., \& {Blinder}, S. 2002, \aap,
  393, 927

\bibitem[{{Benson} \& {Myers}(1989)}]{bm1989}
{Benson}, P.~J., \& {Myers}, P.~C. 1989, \apjs, 71, 89

\bibitem[{{Boss}(1995)}]{boss1995}
{Boss}, A.~P. 1995, in Revista Mexicana de Astronomia y Astrofisica Conference
  Series, Vol.~1, Revista Mexicana de Astronomia y Astrofisica Conference
  Series, ed. {S.~Lizano \& J.~M.~Torrelles}, 165--+

\bibitem[{{Brinch} {et~al.}(2007){Brinch}, {Crapsi}, {J{\o}rgensen},
  {Hogerheijde}, \& {Hill}}]{brinch2007}
{Brinch}, C., {Crapsi}, A., {J{\o}rgensen}, J.~K., {Hogerheijde}, M.~R., \&
  {Hill}, T. 2007, \aap, 475, 915

\bibitem[{{Burkert} \& {Bodenheimer}(1993)}]{burkert1993}
{Burkert}, A., \& {Bodenheimer}, P. 1993, \mnras, 264, 798

\bibitem[{{Burkert} \& {Bodenheimer}(2000)}]{burkert2000}
---. 2000, \apj, 543, 822

\bibitem[{{Caselli} {et~al.}(2002){Caselli}, {Benson}, {Myers}, \&
  {Tafalla}}]{caselli2002}
{Caselli}, P., {Benson}, P.~J., {Myers}, P.~C., \& {Tafalla}, M. 2002, \apj,
  572, 238

\bibitem[{{Cassen} \& {Moosman}(1981)}]{cassen1981}
{Cassen}, P., \& {Moosman}, A. 1981, \icarus, 48, 353

\bibitem[{{Chen} {et~al.}(2007){Chen}, {Launhardt}, \& {Henning}}]{chen2007}
{Chen}, X., {Launhardt}, R., \& {Henning}, T. 2007, \apj, 669, 1058

\bibitem[{{Chevalier}(1983)}]{chevalier1983}
{Chevalier}, R.~A. 1983, \apj, 268, 753

\bibitem[{{Chiang} {et~al.}(2010){Chiang}, {Looney}, {Tobin}, \&
  {Hartmann}}]{chiang2010}
{Chiang}, H., {Looney}, L.~W., {Tobin}, J.~J., \& {Hartmann}, L. 2010, \apj,
  709, 470

\bibitem[{{Di Francesco} {et~al.}(2001){Di Francesco}, {Myers}, {Wilner},
  {Ohashi}, \& {Mardones}}]{difrancesco2001}
{Di Francesco}, J., {Myers}, P.~C., {Wilner}, D.~J., {Ohashi}, N., \&
  {Mardones}, D. 2001, \apj, 562, 770

\bibitem[{{Dib} {et~al.}(2010){Dib}, {Hennebelle}, {Pineda}, {Csengeri},
  {Bontemps}, {Audit}, \& {Goodman}}]{dib2010}
{Dib}, S., {Hennebelle}, P., {Pineda}, J.~E., {Csengeri}, T., {Bontemps}, S.,
  {Audit}, E., \& {Goodman}, A.~A. 2010, \apj, 723, 425

\bibitem[{{Enoch} {et~al.}(2009){Enoch}, {Evans}, {Sargent}, \&
  {Glenn}}]{enoch2009}
{Enoch}, M.~L., {Evans}, N.~J., {Sargent}, A.~I., \& {Glenn}, J. 2009, \apj,
  692, 973

\bibitem[{{Foster} \& {Chevalier}(1993)}]{foster1993}
{Foster}, P.~N., \& {Chevalier}, R.~A. 1993, \apj, 416, 303

\bibitem[{{Froebrich}(2005)}]{froebrich2005}
{Froebrich}, D. 2005, \apjs, 156, 169

\bibitem[{{Goodman} {et~al.}(1993){Goodman}, {Benson}, {Fuller}, \&
  {Myers}}]{goodman1993}
{Goodman}, A.~A., {Benson}, P.~J., {Fuller}, G.~A., \& {Myers}, P.~C. 1993,
  \apj, 406, 528

\bibitem[{{Gueth} {et~al.}(1997){Gueth}, {Guilloteau}, {Dutrey}, \&
  {Bachiller}}]{gueth1997}
{Gueth}, F., {Guilloteau}, S., {Dutrey}, A., \& {Bachiller}, R. 1997, \aap,
  323, 943

\bibitem[{{Hacar} \& {Tafalla}(2011)}]{hacar2011}
{Hacar}, A., \& {Tafalla}, M. 2011, \aap, 533, A34+

\bibitem[{{Hartmann}(2009)}]{hartmann2009}
{Hartmann}, L. 2009, {Accretion Processes in Star Formation: Second Edition},
  ed. {Hartmann, L.} (Cambridge University Press)

\bibitem[{{Hirota} {et~al.}(2011){Hirota}, {Honma}, {Imai}, {Sunada}, {Ueno},
  {Kobayashi}, \& {Kawaguchi}}]{hirota2011}
{Hirota}, T., {Honma}, M., {Imai}, H., {Sunada}, K., {Ueno}, Y., {Kobayashi},
  H., \& {Kawaguchi}, N. 2011, \pasj, 63, 1

\bibitem[{{J{\o}rgensen} {et~al.}(2007){J{\o}rgensen}, {Bourke}, {Myers}, {Di
  Francesco}, {van Dishoeck}, {Lee}, {Ohashi}, {Sch{\"o}ier}, {Takakuwa},
  {Wilner}, \& {Zhang}}]{jorgensen2007}
{J{\o}rgensen}, J.~K., {Bourke}, T.~L., {Myers}, P.~C., {Di Francesco}, J.,
  {van Dishoeck}, E.~F., {Lee}, C., {Ohashi}, N., {Sch{\"o}ier}, F.~L.,
  {Takakuwa}, S., {Wilner}, D.~J., \& {Zhang}, Q. 2007, \apj, 659, 479

\bibitem[{{Kauffmann} {et~al.}(2008){Kauffmann}, {Bertoldi}, {Bourke}, {Evans},
  \& {Lee}}]{kauffmann2008}
{Kauffmann}, J., {Bertoldi}, F., {Bourke}, T.~L., {Evans}, II, N.~J., \& {Lee},
  C.~W. 2008, \aap, 487, 993

\bibitem[{{Kirk} {et~al.}(2009){Kirk}, {Ward-Thompson}, {Di Francesco},
  {Bourke}, {Evans}, {Mer{\'{\i}}n}, {Allen}, {Cieza}, {Dunham}, {Harvey},
  {Huard}, {J{\o}rgensen}, {Miller}, {Noriega-Crespo}, {Peterson}, {Ray}, \&
  {Rebull}}]{kirk2009}
{Kirk}, J.~M., {Ward-Thompson}, D., {Di Francesco}, J., {Bourke}, T.~L.,
  {Evans}, N.~J., {Mer{\'{\i}}n}, B., {Allen}, L.~E., {Cieza}, L.~A., {Dunham},
  M.~M., {Harvey}, P., {Huard}, T., {J{\o}rgensen}, J.~K., {Miller}, J.~F.,
  {Noriega-Crespo}, A., {Peterson}, D., {Ray}, T.~P., \& {Rebull}, L.~M. 2009,
  \apjs, 185, 198

\bibitem[{{Kratter} {et~al.}(2010){Kratter}, {Matzner}, {Krumholz}, \&
  {Klein}}]{kratter2010}
{Kratter}, K.~M., {Matzner}, C.~D., {Krumholz}, M.~R., \& {Klein}, R.~I. 2010,
  \apj, 708, 1585

\bibitem[{{Larson}(1969)}]{larson1969}
{Larson}, R.~B. 1969, \mnras, 145, 271

\bibitem[{{Lee} {et~al.}(2009){Lee}, {Hirano}, {Palau}, {Ho}, {Bourke},
  {Zhang}, \& {Shang}}]{lee2009}
{Lee}, C., {Hirano}, N., {Palau}, A., {Ho}, P.~T.~P., {Bourke}, T.~L., {Zhang},
  Q., \& {Shang}, H. 2009, \apj, 699, 1584

\bibitem[{{Lee} {et~al.}(2004){Lee}, {Bergin}, \& {Evans}}]{lee2004}
{Lee}, J., {Bergin}, E.~A., \& {Evans}, II, N.~J. 2004, \apj, 617, 360

\bibitem[{{Loinard} {et~al.}(2008){Loinard}, {Torres}, {Mioduszewski}, \&
  {Rodr{\'{\i}}guez}}]{loinard2008}
{Loinard}, L., {Torres}, R.~M., {Mioduszewski}, A.~J., \& {Rodr{\'{\i}}guez},
  L.~F. 2008, \apjl, 675, L29

\bibitem[{{Myers}(2005)}]{myers2005}
{Myers}, P.~C. 2005, \apj, 623, 280

\bibitem[{{Myers} {et~al.}(1991){Myers}, {Fuller}, {Goodman}, \&
  {Benson}}]{myers1991}
{Myers}, P.~C., {Fuller}, G.~A., {Goodman}, A.~A., \& {Benson}, P.~J. 1991,
  \apj, 376, 561

\bibitem[{{Narayanan} {et~al.}(2002){Narayanan}, {Moriarty-Schieven}, {Walker},
  \& {Butner}}]{narayanan2002}
{Narayanan}, G., {Moriarty-Schieven}, G., {Walker}, C.~K., \& {Butner}, H.~M.
  2002, \apj, 565, 319

\bibitem[{{Rafikov}(2005)}]{rafikov2005}
{Rafikov}, R.~R. 2005, \apjl, 621, L69

\bibitem[{{Rafikov}(2007)}]{rafikov2007}
---. 2007, \apj, 662, 642

\bibitem[{{Regos} \& {Geller}(1989)}]{regos1989}
{Regos}, E., \& {Geller}, M.~J. 1989, \aj, 98, 755

\bibitem[{{Ryden}(1996)}]{ryden1996}
{Ryden}, B.~S. 1996, \apj, 471, 822

\bibitem[{{Shirley} {et~al.}(2000){Shirley}, {Evans}, {Rawlings}, \&
  {Gregersen}}]{shirley2000}
{Shirley}, Y.~L., {Evans}, II, N.~J., {Rawlings}, J.~M.~C., \& {Gregersen},
  E.~M. 2000, \apjs, 131, 249

\bibitem[{{Shu}(1977)}]{shu1977}
{Shu}, F.~H. 1977, \apj, 214, 488

\bibitem[{{Shu} {et~al.}(1987){Shu}, {Adams}, \& {Lizano}}]{shu1987}
{Shu}, F.~H., {Adams}, F.~C., \& {Lizano}, S. 1987, \araa, 25, 23

\bibitem[{{Smith} {et~al.}(2011){Smith}, {Glover}, {Bonnell}, {Clark}, \&
  {Klessen}}]{smith2011}
{Smith}, R.~J., {Glover}, S.~C.~O., {Bonnell}, I.~A., {Clark}, P.~C., \&
  {Klessen}, R.~S. 2011, \mnras, 411, 1354

\bibitem[{{Smith} {et~al.}(2012){Smith}, {Glover}, {Bonnell}, {Clark}, \&
  {Klessen}}]{smith2012}
{Smith}, R.~J., et al. 2012, submitted

\bibitem[{{Stutz} {et~al.}(2009){Stutz}, {Rieke}, {Bieging}, {Balog},
  {Heitsch}, {Kang}, {Peters}, {Shirley}, \& {Werner}}]{stutz2009}
{Stutz}, A.~M., {Rieke}, G.~H., {Bieging}, J.~H., {Balog}, Z., {Heitsch}, F.,
  {Kang}, M., {Peters}, W.~L., {Shirley}, Y.~L., \& {Werner}, M.~W. 2009, \apj,
  707, 137

\bibitem[{{Tafalla} {et~al.}(1998){Tafalla}, {Mardones}, {Myers}, {Caselli},
  {Bachiller}, \& {Benson}}]{tafalla1998}
{Tafalla}, M., {Mardones}, D., {Myers}, P.~C., {Caselli}, P., {Bachiller}, R.,
  \& {Benson}, P.~J. 1998, \apj, 504, 900

\bibitem[{{Tanner} \& {Arce}(2011)}]{tanner2011}
{Tanner}, J.~D., \& {Arce}, H.~G. 2011, \apj, 726, 40

\bibitem[{{Terebey} {et~al.}(1984){Terebey}, {Shu}, \& {Cassen}}]{tsc1984}
{Terebey}, S., {Shu}, F.~H., \& {Cassen}, P. 1984, \apj, 286, 529

\bibitem[{{Tobin} {et~al.}(2011){Tobin}, {Hartmann}, {Chiang}, {Looney},
  {Bergin}, {Chandler}, {Masqu{\'e}}, {Maret}, \& {Heitsch}}]{tobin2011}
{Tobin}, J.~J., {Hartmann}, L., {Chiang}, H.-F., {Looney}, L.~W., {Bergin},
  E.~A., {Chandler}, C.~J., {Masqu{\'e}}, J.~M., {Maret}, S., \& {Heitsch}, F.
  2011, \apj, 740, 45

\bibitem[{{Tobin} {et~al.}(2010){Tobin}, {Hartmann}, {Looney}, \&
  {Chiang}}]{tobin2010a}
{Tobin}, J.~J., {Hartmann}, L., {Looney}, L.~W., \& {Chiang}, H. 2010, \apj,
  712, 1010

\bibitem[{{Ulrich}(1976)}]{ulrich1976}
{Ulrich}, R.~K. 1976, \apj, 210, 377

\bibitem[{{Visser} {et~al.}(2002){Visser}, {Richer}, \&
  {Chandler}}]{visser2002}
{Visser}, A.~E., {Richer}, J.~S., \& {Chandler}, C.~J. 2002, \aj, 124, 2756

\bibitem[{{Volgenau} {et~al.}(2006){Volgenau}, {Mundy}, {Looney}, \&
  {Welch}}]{volgenau2006}
{Volgenau}, N.~H., {Mundy}, L.~G., {Looney}, L.~W., \& {Welch}, W.~J. 2006,
  \apj, 651, 301

\bibitem[{{Walker} {et~al.}(1986){Walker}, {Lada}, {Young}, {Maloney}, \&
  {Wilking}}]{walker1987}
{Walker}, C.~K., {Lada}, C.~J., {Young}, E.~T., {Maloney}, P.~R., \& {Wilking},
  B.~A. 1986, \apjl, 309, L47

\bibitem[{{Ward-Thompson} \& {Buckley}(2001)}]{wtbuckley2001}
{Ward-Thompson}, D., \& {Buckley}, H.~D. 2001, \mnras, 327, 955

\bibitem[{{Young} {et~al.}(2006){Young}, {Enoch}, {Evans}, {Glenn}, {Sargent},
  {Huard}, {Aguirre}, {Golwala}, {Haig}, {Harvey}, {Laurent}, {Mauskopf}, \&
  {Sayers}}]{young2006}
{Young}, K.~E., {Enoch}, M.~L., {Evans}, II, N.~J., {Glenn}, J., {Sargent}, A.,
  {Huard}, T.~L., {Aguirre}, J., {Golwala}, S., {Haig}, D., {Harvey}, P.,
  {Laurent}, G., {Mauskopf}, P., \& {Sayers}, J. 2006, \apj, 644, 326

\bibitem[{{Zhou}(1992)}]{zhou1992}
{Zhou}, S. 1992, \apj, 394, 204

\bibitem[{{Zhou} {et~al.}(1993){Zhou}, {Evans}, {Koempe}, \&
  {Walmsley}}]{zhou1993}
{Zhou}, S., {Evans}, II, N.~J., {Koempe}, C., \& {Walmsley}, C.~M. 1993, \apj,
  404, 232

\bibitem[{{Zhu} {et~al.}(2010){Zhu}, {Hartmann}, \& {Gammie}}]{zhu2010}
{Zhu}, Z., {Hartmann}, L., \& {Gammie}, C. 2010, \apj, 713, 1143

\end{thebibliography}
